\documentclass[preprint,12pt]{revtex4-1}
\usepackage{amsmath}
\usepackage{amssymb}
\usepackage{amscd}
\usepackage{pstricks}
\usepackage{pst-all}
\usepackage[all]{xy}
    \SelectTips{cm}{}
\usepackage[mathscr]{euscript}
\usepackage{graphicx}
\usepackage{booktabs}
\usepackage{theorem}
\theoremstyle{plain}
\newtheorem{thm}{Theorem}
\newtheorem{prp}[thm]{Proposition}
\newtheorem{dfn}[thm]{Definition}

\begin{document}

\title{Sheaves in Quantum Topos Induced by Quantization}
\author{Kunji Nakayama\footnote{e-mail: nakayama@law.ryukoku.ac.jp}
\\
Faculty of Law\\
Ryukoku University\\
Fushimi-ku, Kyoto 612-8577}

\maketitle

\def\Rmath{{\mathbb{R}}}

\def\Set{{\bf Set}}
\def\Sets{{\bf Sets}}
\def\Sh{{\rm Sh}}
\def\op{{\rm op}}

\def\Ocal{{\cal O}}
\def\Ocalop{{\cal O}^{\rm op}}
\def\OcalBorelR{\Ocal \times \mathfrak{B}_{\mathbb{R}}}
\def\Obj{{\rm Obj}}
\def\Mor{{\rm Mor}}

\def\Ahat{\hat{A}}
\def\Bhat{\hat{B}}
\def\Ehat{\hat{E}}
\def\Hhat{\hat{H}}
\def\Rhat{\hat{R}}

\def\Aboldhat{\widehat {\bf A}}
\def\Eboldhat{\widehat {\bf E}}
\def\Cboldhat{\widehat {\bf C}}
\def\Pboldhat{\widehat {\bf P}}
\def\Vboldhat{\widehat {\bf V}}
\def\Vcalhat{\widehat {\cal V}}

\def\Aboldop{{\bf A^{{\rm op}}}}
\def\Cboldop{{\bf C^{{\rm op}}}}
\def\Eboldop{{\bf E^{{\rm op}}}}
\def\Vboldop{{\bf V^{{\rm op}}}}
\def\Vcalop{{\cal V^{{\rm op}}}}

\def\Aboldop{{\bf A}^{\rm op}}

\def\Vboldupsilon{{\bf V}_{\uptilde}}
\def\Vboldupsilonop{{\bf V}_{\uptilde}^{\rm op}}
\def\Vboldupsilonhat{\hat{{\bf V}_{\uptilde}}}

\def\CboldOcal{{\bf C}_{{\cal O}}}
\def\CboldOcalop{{\bf C}_{{\cal O}}^{\rm op}}
\def\CboldOcalhat{\widehat{{\bf C}_{{\cal O}}}}

\def\alphabold{{\bf \alpha}}

\def\ybold{{\bf y}}
\def\Abold{{\bf A}}

\def\Bbold{{\bf B}}
\def\Cbold{{\bf C}}
\def\Gbold{{\bf G}}
\def\Ebold{{\bf E}}
\def\Fbold{{\bf F}}
\def\Mbold{{\bf M}}
\def\Nbold{{\bf N}}
\def\Kbold{{\bf K}}
\def\Hbold{{\bf H}}
\def\Lbold{{\bf L}}
\def\Pbold{{\bf P}}
\def\Vbold{{\bf V}}
\def\Sbold{{\bf S}}
\def\Tbold{{\bf T}}
\def\Omegabold{{\bf \Omega}}

\def\Kcal{{\cal K}}

\def\Pcal{{\cal P}}
\def\Qcal{{\cal Q}}
\def\betatilde{\tilde{\beta}}

\def\Subsets{{\rm Subsets}}

\def\Hom{{\rm Hom}}
\def\Sub{{\rm Sub}}
\def\dom{{\rm dom}}
\def\cod{{\rm cod}}

\def\Flt{{\cal F}}
\def\Dcal{{\mathfrak D}}
\def\Ecal{{\mathfrak E}}
\def\Fcal{{\mathfrak F}}
\def\Lcal{{\mathfrak L}}
\def\Ocal{{\cal O}}
\def\Scal{{\cal S}}

\def\Amath{{\mathfrak A}}
\def\Bmath{{\mathfrak B}}
\def\Cmath{{\mathfrak C}}
\def\Emath{{\mathfrak E}}

\def\Scalbar{{\bar{\cal S}}}
\def\Hcal{{\cal H}}
\def\Ccal{{\cal C}}
\def\Gcal{{\cal G}}
\def\Rcal{{\cal R}}

\def\Tcal{{\cal T}}
\def\Vcal{{\cal V}}
\def\tmath{{\mathfrak{t}}}
\def\Dmath{{\mathfrak D}}
\def\Fmath{{\mathfrak F}}
\def\Vmath{{\mathfrak V}}

\def\CcalR{{{\cal C}_R}}
\def\CcalRe{{{\cal C}_R^{e \downarrow}}}
\def\Ocaltilde{{\cal O}/_{\sim}}

\def\ahat{\hat{a}}
\def\Fhat{\hat{F}}
\def\Hhat{\hat{H}}
\def\Ihat{\hat{I}}

\def\Ghat{\hat{G}}

\def\Com{{\rm Com}}
\def\ES{{\rm ES}}
\def\op{{\rm op}}
\def\Onebold{{\bf 1}}
\def\Mor{{\rm Mor}}
\def\Obj{{\rm Obj}}
\def\Bcal{{\cal B}}
\def\End{{\rm End}}

\def\true{{\rm true}}
\def\onepoint{\{ \, \cdot \,\}}
\def\id{{\rm id}}
\def\pt{{\rm pt}}

\def\alphabreve{\breve{\alpha}}
\def\uptilde{\tilde{\upsilon}}

\def\vertin{\mathop{{\rotatebox{90}{$\in$}}}}

This paper shows that quantization induces 
a Lawvere-Tierney topology on (hence, a sheaf topos in) the quantum topos.
We show that a quantization map from classical observables to
self-adjoint operators on a Hilbert space naturally induces geometric morphisms 
from presheaf topoi related to the classical system into
a presheaf topos, called the quantum topos, 
on the context category consisting of commutative von Neumann algebras
of bounded operators on the Hilbert space.
By means of the geometric morphisms, 
we define Lawvere-Tierney topologies on the quantum topos
(and their equivalent Grothendieck topologies on the context category). 
We show that, among them, we can uniquely select the most informative one, 
which we call the quantization topology.
We furthermore construct sheaves induced by the quantization topology.
This can be done in an elementary and self-contained way,
because the quantization topology has a quite simple expression.


\section{Introduction}
\label{sec:Introduction}

For more than a decade, 
Isham and his collaborators has challenged to
reformulate and develop quantum theory 
on the topos theoretic base~\cite{IB98,BI99,HBI00,BI02,DI08a,DI08b,DI08c,DI08d}.
In their approach,
quantum theory is coded in 
the presheaf topos $\Set^{\Vbold(\Hcal)^{\rm op}}$,
where $\Vbold(\Hcal)$ is the category 
of commutative von Neumann algebra of bounded operators
on a given Hilbert space $\Hcal$ 
with set inclusions as morphisms.
By means of the novel mathematical framework,
they succeeded in liberating
quantum theory from the instrumentalist interpretation;
quantum theory can be reformulated
in a way that is fundamentally realistic and objective independently of 
the notion of measurement. 
Topos quantum theory, therefore, promises to give a consistent framework
for quantum gravity and quantum cosmology where 
the instrumentalist interpretation cannot be consistent.

The topos quantum theory is developed 
as a general theory of quantum systems from the beginning;
it does not need classical systems corresponding to quantum systems,
as is the case for, e.g., the axiomatic quantum theory 
based on the Hilbert space theory.
When we have a classical system corresponding to a quantum system,
however,
we need quantization of the classical system
in order to obtain its quantum theory on $\Set^{\Vbold(\Hcal)^{\rm op}}$,
as is the case with the standard formulation of quantum physics.
Via the quantization,
some operators on the Hilbert space $\Hcal$ are related to
classical observables.
And then, we can expect that 
the correspondence between classical observables
and operators on $\Hcal$
induces some substructure in the quantum topos.

If there exists a classical system corresponding to a quantum system,
what structure does the quantization induce 
in the quantum topos?
This is our primary question in the present paper.
It includes also the following question;
how is the structure of the classical system reflected 
on the quantum topos? 
Or, how is the former characterized in the latter?
Furthermore, 
the author thinks that it could be meaningful to ask 
the following simple question:
What is `quantization' for the topos quantum theory?
Or, how should the notion of quantization be generalized to
the topos theoretic language?
The topos quantum theory might actually discard Hilbert spaces
in development.
If that is the case and if 
it is still meaningful to consider quantization of classical systems,
the notion of quantization has to be described
in terms of topos theoretic notions independent of those of Hilbert spaces.
It is, therefore, desirable to obtain a guiding sample for the question.
This is our underlying motivation in the present paper.

We formulate our question as follows;
Suppose that we are given classical observables $\Ocal$ to be quantized,
a Hilbert space $\Hcal$,
and a quantization map $\upsilon$ which assigns faithfully 
a self-adjoint operator on $\Hcal$
to each classical observable.
Then, what structures are induced to 
the quantum presheaf topos $\Set^{\Vbold(\Hcal)^{\rm op}}$?
This question could be answered variously.
Our answer in this paper is, however, quite simple and clear:
Quantization induces a Lawvere-Tierney topology on $\Set^{\Vbold(\Hcal)^{\rm op}}$
(hence, a Grothendieck topology on $\Vbold(\Hcal)$,
which we call the quantization topology later;
We can, therefore, define a topos,
the quantization topos which consists of sheaves induced by the quantization topology.

This paper is organized as follows.
In \S\ref{sec:Prequantization Categories},
we define the notion of prequantization categories $\Cbold$
related to the set $\Ocal$ of observables of classical theory.
We show that a given quantization map naturally induce 
a functor, which we call the quantization functor, from $\Cbold$ to $\Vbold(\Hcal)$.
Also, the key prequantization category,
which we call the proper prequantization category,
is derived.
In \S\ref{sec:Geometric Morphisms},  
we define prequantization topoi at the beginning.
We construct geometric morphisms
from them to the quantum topos. 
The morphisms are induced by the quantization functors.
In \S\ref{sec:Topologies on Quantum},
we construct Lawvere-Tierney topologies and 
Grothendieck topologies which the geometric morphisms
induce on the quantum topos.
Among them,
we, further, select a particular topology
which we call the quantization topology.
It is induced by any 
prequantization category which includes the proper prequantization 
category as a subcategory.
In \S\ref{sec:Quantization Sheaf Topos}, we give a condition 
by which a presheaf in the quantum topos 
is a sheaf for the quantization topology.
Also, the associated sheaf functor is derived.
Furthermore, we show that the category of the quantization sheaves are indeed a topos.
In \S\ref{sec:Prequantization Topoi and}, 
we show interrelationships among the prequantization topoi
and the quantization sheaf topos.
In \S\ref{sec:Concluding Remarks}, we make some remarks.

\section{Prequantization Categories}
\label{sec:Prequantization Categories}

\subsection{Galois connection induced by quantization map}
\label{subsec:Galois connection induced}

A set of classical observables $\Ocal$ to be quantized is 
a Lie algebra with respect to appropriately defined commutator,
such as a Poisson algebra of a set of functions on phase space.
Quantization of $\Ocal$ is given by an irreducible map
\begin{equation}
\upsilon: a \mapsto \ahat,
\label{eq:quantizationmap1}
\end{equation}
which assigns a self-adjoint operator $\ahat$ on a Hilbert space $\Hcal$
to each element $a \in \Ocal$
in such a way that the Lie-noncommutativity in $\Ocal$
is reflected by the noncommutativity in the operator algebra on $\Hcal$.
We assume that $\Ocal$,
$\upsilon$, and $\Hcal$ are given from the beginning.
(Therefore, we abbreviate $\Vbold(\Hcal)$ to $\Vbold$ hereafter.)
Also, we suppose that the quantization map $\upsilon$ is faithful;
that is, for all $a$, $b \in \Ocal$,
\begin{equation}
a \neq b \implies \ahat \neq \hat{b}.
\label{eq:quantizationfaithful}
\end{equation}
In general, $\ahat$ is not always a bounded operator.
In order to relate each classical observable to 
a bounded operator on $\Hcal$,
we define a map $\uptilde$, also which we call the quantization map, by
\begin{equation}
\uptilde:a \mapsto \ahat \mapsto e^{i \ahat}.
\label{eq:quantizationmap2}
\end{equation}

Let $\CboldOcal$ be a collection of all subsets $C$ of $\Ocal$ 
such that,
for any $a$, $b \in C$, $[a,b] = 0$.
It is obvious that $\CboldOcal$ is a category
whose morphisms are set inclusions.
We give the following definition:
\begin{dfn}
We call full subcategories of $\CboldOcal$ prequantization categories.
\end{dfn}
If any object of a prequantization category $\Cbold'$ is
also that of another one $\Cbold$,
there exists an inclusion functor $\Cbold' \hookrightarrow \Cbold$.
Therefore, the collection of all prequantization categories is itself 
a category whose morphisms are inclusion functors.
We write $\Cmath$ for it.

We can construct a functor $\phi:\CboldOcal \to \Vbold$
from the quantization map $\uptilde$ as follows:
Since $C \in \CboldOcal$ is a set of Lie-commutative classical observables,
$\Upsilon(C) := \uptilde(C) \cup \uptilde(C)^{*}$ is a set of commutative unitary operators.
We define $\phi(C)$ as the smallest commutative von Neumann algebra
which includes $\Upsilon(C)$;
namely,
\begin{equation}
\phi(C) := \Upsilon(C)' \cap \Upsilon(C)'' = \Upsilon(C)'',
\end{equation} 
where $'$ is the commutant operator.
Since  $\phi(C') \subseteq \phi(C)$ whenever $C' \subseteq C$,
$\phi$ is a functor from $\CboldOcal$ to $\Vbold$.
Also, for each $\Cbold \in \Cmath$,
we can define a functor $\phi|_{\Cbold}: \Cbold \to \Vbold$
as the restriction of $\phi$ on $\Cbold$.

We call the functor $\phi: \CboldOcal \to \Vbold$ 
and its restrictions $\phi|_{\Cbold}$ on subcategories $\Cbold$ {\it quantization functors}
induced by the quantization map $\uptilde$.
(We abbreviate the symbol $\phi|_{\Cbold}$ as $\phi$ hereafter.)
Obviously, 
each inclusion $\Cbold' \hookrightarrow \Cbold \in \Cmath$
makes the diagram
\begin{equation}
\xymatrix{
\Cbold' \ar @{^{(}->} [rr] \ar [rdd] _{\phi}  && \Cbold \ar [ldd] ^{\phi} \\
&& \\
&\Vbold &  \\
}
\label{eq:C'CV}
\end{equation}
commute.

Next, we define a functor $\psi:\Vbold \to \CboldOcal$ by
\begin{equation}
\psi(V) 
 := 
\uptilde^{-1}(V)
 = 
\{ 
a \in \Ocal \,|\, \uptilde(a) \in V\}
\label{eq:IQF}
\end{equation}
for every $V \in \Vbold$.
Since $\uptilde$ is assumed to be faithful, 
$\psi(V)$ consists of Lie-commutative sets of classical observables,
that is, $\psi(V) \in \CboldOcal$.
Also, it is obvious that $\psi$ preserves inclusions.
Therefore, we can regard $\psi$ as a functor from $\Vbold$ into $\CboldOcal$.
We such a functor $\psi:\Vbold \to \CboldOcal$
a {\it classicization functor} induced by $\uptilde$.

We have the following proposition on $\phi$ and $\psi$.
\begin{prp}
The pair $(\phi, \psi)$ of the quantization functor and 
the classicization functor  is a Galois connection.
That is,
for every $C \in \CboldOcal$ and $V \in \Vbold$, 
the equivalence relation
\begin{equation}
\phi(C) \subseteq  V
\iff
C \subseteq \psi (V)
\label{eq:Galois}
\end{equation}
holds.
Or equivalently, it follows that
\begin{equation}
C \subseteq \sharp (C) ,
\label{eq:1<sharp}
\end{equation}
and
\begin{equation}
\flat (V) \subseteq V,
\label{eq:flat<1}
\end{equation}
where the endofunctors
$\sharp:\CboldOcal \to \CboldOcal$ and $\flat:\Vbold \to \Vbold$ are 
defined by
\begin{equation}
\sharp := \psi \circ \phi 
\quad
{\it and}
\quad
\flat := \phi \circ \psi.
\label{eq:sharpflat}
\end{equation}
\end{prp}
{\it Proof.}
Inclusion relations (\ref{eq:1<sharp}) and (\ref{eq:flat<1}) are obvious.
In fact, since $\phi (C)$ is a commutative von Neumann algebra
which includes $\uptilde (C)$, its inverse contains $\psi(\phi(C))$
includes $C$.
Also, since $V$ is a commutative von Neumann algebra 
including $\uptilde(\uptilde^{-1}(V))=\uptilde(\psi(V))$,
it includes $\phi(\psi(V))$.

In order to verify equivalence relation (\ref{eq:Galois}),
assume that $\phi(C) \subseteq V$.
Then, it follows that
\begin{equation}
\psi( \phi(C)) = \sharp (C) \subseteq \psi(V).
\end{equation}
Therefore, $C \subseteq \psi(V)$ follows from (\ref{eq:1<sharp}).
Conversely, if $C \subseteq \psi(V)$ holds, then we have
\begin{equation}
\phi(C) \subseteq  \phi (\psi (V)) = \flat (V) ,
\end{equation}
hence, $\phi(C) \subseteq  V$
follows from (\ref{eq:flat<1}).

Note that we derived the Galois connection relation (\ref{eq:Galois})
from relations (\ref{eq:1<sharp}) and (\ref{eq:flat<1}).
We should note, however,  that if  (\ref{eq:Galois}) holds,
(\ref{eq:1<sharp}) and (\ref{eq:flat<1}) follow from it.
That is, they are equivalent.
In fact, since $\phi(C) \subseteq \phi(C)$,
we have (\ref{eq:1<sharp}) by (\ref{eq:Galois}).
Also, $\psi(V) \subseteq \psi(V)$ gives  (\ref{eq:flat<1}).
\hspace{\fill}$ \square $

For later convenience, 
we should mention a few equalities which are satisfied by any Galois connection pair
$(\phi, \psi)$~\cite{DP90}:
The functors $\phi:\CboldOcal \to \Vbold$ and
$\psi:\Vbold \to \CboldOcal$ satisfy equalities
\begin{equation}
\phi \psi \phi = \phi
\quad
\mbox{and}
\quad
\psi \phi \psi = \psi.
\label{eq:ppp=p}
\end{equation}
In fact, for every $C \in \CboldOcal$, we have $C \subseteq \sharp (C)$,
hence $\phi(C) \subseteq \phi\psi\phi(C)$,
whereas $\phi\psi\phi(C) =\flat(\phi(C)) \subseteq \phi(C)$.
Thus, $\phi\psi\phi(C) = \phi(C)$.
Similarly, $\psi \phi \psi (V) = \psi (V)$ holds for
every $V \in \Vbold$.
Equations (\ref{eq:ppp=p}) imply that
the endofunctors $\sharp$ and $\flat$ are idempotents:
\begin{equation}
\sharp\sharp= \sharp
\quad
\mbox{and}
\quad
\flat\flat=\flat.
\label{eq:sssfff}
\end{equation}

\subsection{Proper prequantization category}
\label{subsec:Proper prequantization category}

So far, the objects of $\Cbold$ have not been assumed to have
algebraic structures.
Some prequantization categories, however, 
have objects equipped with some algebraic structures.
Among them, 
there exists a distinct one,
the {\it proper prequantization category} $\Abold$,
which plays a key role in the following.
It is defined as the collection of all fixpoints of $\sharp \equiv \psi \circ \phi$;
namely, for every $C \in \CboldOcal$,
\begin{equation}
C \in \Abold
\quad
\iff
\quad
\sharp(C) = C.
\label{eq:PPC}
\end{equation}
It is easy to see that the right hand side of condition (\ref{eq:PPC}) 
is equivalent to the condition that
\begin{equation}
\exists V \in \Vbold \mbox{ s.t. } C = \psi(V).
\label{PPC'}
\end{equation}
In fact, if $C = \sharp(C)$ holds for $C \in \CboldOcal$, 
we can take $V=\phi(C) \in \Vbold$.
Conversely, if we have $V \in \Vbold$ such that $C= \psi(V)$,
then it follows that
\begin{equation}
C \subseteq \sharp(C) = \psi\circ \phi ( \psi(V)) = \psi(V)= C.
\end{equation}
\begin{prp}
Each object $A \in \Abold$ is a Lie-commutative algebra.
\end{prp}
{\it Proof.}
Since $\Ocal$ contains $0$ as a Lie algebra
and $\phi(A)$ contains a unit $I= \uptilde(0)$ as a von Neumann algebra, 
$\psi(I) = 0 \in \psi(\phi(A))$.
The commutative set $A$ is, therefore,
closed for the Lie-bracket of $\Ocal$.
If $a$, $b \in A$, 
then $\uptilde(a)$, $\uptilde(b) \in \phi(A)$,
hence $\uptilde(a+b)=\uptilde(a)\uptilde(b) \in \phi(A)$.
Therefore, $a+b=b+a \in \psi(\phi(A))= \sharp(A)=A$,
since $a+b = b+a \in \Ocal$.
Also, if $a \in A$, then $\uptilde(a) = \exp(i \hat{a})\in \phi(A)$,
while for any $k \in \mathbb{R}$, $ka \in \Ocal$.
Since $\uptilde(ka)=\exp(i k \hat{a})=f(\exp(a))$
where $f(x)=x^{k}$ defined on the spectral set of $\exp(a)$,
$\phi(A)$ contains $\uptilde(ka)$ as a $C^{*}$-algebra, hence, $ka \in \Abold$.
Thus, $A \in \Abold$ is a Lie-commutative algebra.
\hspace{\fill}$\square$

Note that each morphism of $\Abold$ is a set inclusion,
hence a homomorphism of algebra.
Also, each $A \in \Abold$ 
can be regarded as a topological Lie-commutative algebra
by transferring topologies on the von Neumann algebra $\phi(A)$
by the map $\uptilde^{-1}$.
The present paper, however,
deals with neither the algebraic properties 
nor the topological properties of objects of $\Abold$.

Returning to the category theoretic properties on 
the proper prequantization category $\Abold$,
we consider the endofunctor $\sharp:\CboldOcal \to \CboldOcal$.
Since, for each $C \in \CboldOcal$, $\sharp(C) \in \Abold$
from the definition of $\Abold$,
we can restrict the codomain of $\sharp$ to any prequantization category
which includes $\Abold$ as a subcategory.
Also, of course, the domain can be restricted to arbitrary prequantization categories.
We write $\sharp$ for also the restricted functors of $\sharp$.

We note that the diagram
\begin{equation}
\xymatrix{
\Abold \ar @{^{(}->} [rr] \ar @{=} [dd] && \Cbold \ar [lldd] ^{\sharp}\\
&& \\
\Abold &&  \\
}
\label{eq:ACA}
\end{equation}
commutes because of definition (\ref{eq:PPC}) of $\Abold$.
Here,  $\Cbold$ is an arbitrary prequantization category including $\Abold$.
Furthermore, note that the left-hand equality on (\ref{eq:ppp=p})
implies that, for any $\Cbold \in \Cmath$,
its quantization functor $\phi:\Cbold \to \Vbold$
can be factored through $\Abold$.
That is, the diagram

\begin{equation}
\xymatrix{
\Cbold \ar [rr] ^{\sharp} \ar [rdd] _{\phi} && \Abold \ar [ldd] ^{\phi}  \\
&& \\
& \Vbold  & \\
}
\label{eq:CAV}
\end{equation}
commutes.
Summing up the commutative diagrams (\ref{eq:ACA}) and (\ref{eq:CAV})
with equality (\ref{eq:ppp=p}), 
we obtain the following proposition:
\begin{prp}
For any prequantization category $\Cbold \in \Cmath$
which includes $\Abold$ as a subcategory
and for any $\Cbold' \in \Cmath$,
the diagram
\begin{equation}
\xymatrix{
\Cbold' \ar [rr] ^{\sharp} \ar [rrdd] _-{\phi } && \Abold \ar [dd] ^-{\phi} \ar @{^{(}->} [rr]  && \Cbold \ar [lldd] ^-{\phi} \\
&& && \\
&&\Vbold && \\
}
\end{equation}
commutes.
\end{prp}
The proper prequantization category $\Abold$ is, therefore,
the least one among the prequantization categories
through which every quantization functor can be factored.

We should note that,
corresponding to $\Abold$, a subcategory $\Vbold_{\flat}$ of $\Vbold$
can be defined as a collection of fixpoints of $\flat$:
For all $V \in \Vbold$
\begin{eqnarray}
V \in \Vbold_{\flat}
& \iff &
\flat(V) = V \\
& \iff &
\exists C\in \CboldOcal
\quad{\rm s.t.}\quad
V = \phi(C)
\label{eq:Vflat}
\end{eqnarray}
It is obvious that $\phi$ and $\psi|_{\Vbold_{\flat}}$
give order isomorphisms between $\Abold$
and $\Vbold_{\flat}$.

\subsection{Invariance of Galois connection $(\phi,\psi)$ under parametrized quantization maps}
\label{subsec:Invariance of the Galois connection}

Finally, we should note that we have freedom 
to define a quantization map $\uptilde$.
That is, for a given (\ref{eq:quantizationmap1}), 
we could define a quantization map as
\begin{equation}
\uptilde_{k}: a \mapsto \hat{a} \mapsto {\rm e}^{i k\hat{a}}
\label{eq:quantizationmap3}
\end{equation}
with $k \in \Rmath \setminus \{0\}$,
instead of (\ref{eq:quantizationmap2}).
Correspondingly,
we could have a quantization functor $\phi_{k}$,
the corresponding classicization functor $\psi_{k}$, 
and the proper prequantization category $\Abold_{k}$.
The functors $\phi$ and $\psi$
and the proper prequantization category $\Abold$ 
are, therefore, the special case where $k = 1$.
We can show, however, the following fact:
\begin{prp}
Quantization functors and classicization functors are invariant
under the transformation $\uptilde \mapsto \uptilde_{k}$;
namely, 
\begin{equation}
\phi_{k} = \phi
\quad \mbox{and} \quad
\psi_{k} = \psi,
\label{eq:ppinvariance}
\end{equation}
hence, for the proper prequantization categories,
\begin{equation}
\Abold_{k} = \Abold.
\label{eq:Ainvariance}
\end{equation}
\end{prp}
{\it Proof.}
Let $C$ be a prequantization category 
and $a \in C$.
Then, $\uptilde(a) \in \phi(C)$ and $\uptilde_{k}(a) \in \phi_{k}(C)$.
We have, however,
\begin{equation}
\uptilde_{k}(a) = f(\uptilde(a))
\quad
\mbox{and}
\quad
\uptilde(a) = g(\uptilde_{k}(a)) ,
\end{equation}
where $f$ (resp. $g$) are functions defined on
the spectrum of $\uptilde(a)$ (resp. $\uptilde_{k}(a)$)
defined by
$f(x) = x^{k}$ and $g(x) = x^{1/k}$.
Therefore, 
it follows that, for all $a \in C$,
\begin{equation}
\uptilde_{k}(a) \in \phi(C)  
\quad
\mbox{and}
\quad
\uptilde(a) \in \phi_{k}(C),
\label{eq:utainphi-uainphit}
\end{equation}
since $\phi(C)$ and $\phi_{k}(C)$
are $C^{*}$-algebras as von Neumann algebras. 
It is obvious that equation (\ref{eq:utainphi-uainphit}) implies $\phi_{k}(C) \subseteq \phi(C)$
and $\phi(C) \subseteq \phi_{k}(C)$.
Thus, we can conclude $\phi_{k}(C) = \phi(C)$ for every $C \in \Cbold(\Ocal)$,
which implies (\ref{eq:ppinvariance}),
and hence (\ref{eq:Ainvariance}).
\hspace{\fill}$\square$

\section{Geometric Morphisms}
\label{sec:Geometric Morphisms}

First, we give the following definition:
\begin{dfn}
We call a topos $\Set^{\Cboldop}$ of presheaves 
on a prequantization category $\Cbold \in \Cmath$ a prequantization topos.
\end{dfn}

In the previous subsection,
we constructed the functors from prequantization categories $\Cbold$
of classical observables to the category $\Vbold$ of commutative von Neumann algebras
of bouded operators on a Hilbert space $\Hcal$.
According to the general theory of topoi, 
any such functor naturally induces a geometric morphism from
the corresponding prequantization topos to the quantum topos~\cite{MM92}.
Here, a geometric morphism $\phi : \Set^{\Cboldop} \to \Set^{\Vboldop}$ is a pair of functors
\begin{equation}
\xymatrix{
\Set^{\Cboldop} \ar @<-1mm>[r] _{\phi_{*}} & \Set^{\Vboldop} \ar @<-1mm> [l] _{\phi^{*}} \\
}
\end{equation}
which satisfies the following conditions:
\begin{itemize}
\item[(G1)] The functor $\phi^{*}$ is left adjoint to $\phi_{*}$;
namely, for each $P \in \Set^{\Cboldop} $ and $Q \in \Set^{\Vboldop}$, 
there exists a bijection 
\begin{equation}
\Hom_{\Cboldhat}(\phi^{\ast}(Q),P) \simeq \Hom_{\Vboldhat}(Q,\phi_{\ast}(P)) 
\end{equation}
which is natural for $P$ and $Q$.
\item[(G2)] The functor $\phi^{*}$ is left exact; that is, it preserves any finite limit.
\end{itemize}
Here, $\phi_{*}$ is called the direct image part of $\phi$,
and $\phi^{*}$ the inverse image part.
Also, $\Cboldhat$ and $\Vboldhat$,
which we often use hereafter, are abbreviations for $\Set^{\Cboldop}$
and $\Set^{\Vboldop}$, respectively.

The purpose of this section is to give explicit expressions of
$\phi^{*}$ and $\phi_{*}$.
They can be derived from the well-known generalized expressions~\cite{MM92}.
In Appendix \ref{subsec:Adjunction},
we prove that they indeed satisfy condition (G1).
It is well-known that, in general, 
if a geometric morphism between presheaf topoi is induced by
a functor between their base categories,
(G2) is ensured automatically 
by the existence of a left-adjoint to the inverse image part.
In Appendix \ref{sec:Adjunction}, we construct a functor $\phi !$ 
which is left adjoint to $\phi^{*}$ as a proof of (G2).


The inverse image $\phi^{\ast}:\Set^{\Vboldop} \to \Set^{\Cboldop}$ is defined as follows:
For any object $Q \in \Set^{\Vboldop}$, $\phi^{\ast}(Q)\in \Set^{\Cboldop}$
is defined by
\begin{equation}
\phi^{\ast} Q \equiv Q_{\phi} := Q \circ \phi^{\op}.
\label{eq:inverseobj}
\end{equation}

More  precisely,
for each $C \in \Cbold$,
\begin{equation}
\phi^{\ast}Q(C)  \equiv Q_{\phi}(C) := Q(\phi(C)),
\label{eq:inverseobjobj}
\end{equation}
and, for each $C' \hookrightarrow C \in \Mor(\Set^{\Cboldop})$ and $q^{\phi} \in \phi^{*}Q(C)$,
\begin{eqnarray}
\phi^{\ast}Q(C' \hookrightarrow C) (q^{\phi}) 
& \equiv &
Q_{\phi}(C' \hookrightarrow C) (q^{\phi}) \equiv q^{\phi}||_{C'}\nonumber\\
& := & 
Q(\phi(C') \hookrightarrow \phi(C)) (q^{\phi}) 
\equiv
q^{\phi}|_{\phi(C')}.
\label{eq:inverseobjobj}
\end{eqnarray}

For each morphism $\xymatrix{ Q \ar [r] ^{\theta} & Q'} \in \Mor(\Set^{\Vboldop})$,
the corresponding morphism $\xymatrix{ \phi^{\ast}Q \ar [r] ^{\phi^{\ast}\theta} & \phi^{\ast}Q'} \in \Mor(\Set^{\Cboldop})$ 
is defined by
\begin{equation}
(\phi^{*}\alpha)_{C}(q^{\phi})
 \equiv 
(\theta_{\phi})_{C}(q^{\phi}) 
 := 
(\theta_{\phi (C)})(q^{\phi}) \in Q'(\phi(C)) = \phi^{*}Q'(C)
\end{equation}
for each $C \in \Cbold$ and $q^{\phi} \in \phi^{*}Q(C)= Q(\phi(C))$.
It is easy to see that $\xymatrix{ \phi^{\ast}Q \ar [r] ^{\phi^{\ast}\theta} & \phi^{\ast}Q'}$
is indeed a morphism in $\Set^{\Cboldop}$;
the naturality condition for $\phi^{*}\theta$,
that is, the commutativity of the diagram 
\begin{equation}
\xymatrix{
\phi^{*}Q(C)  \ar [rr] ^{\phi^{*} \theta_{C}} \ar [dd] _{\phi^{*}Q (C' \hookrightarrow C)} & & \phi^{*}Q' (C) \ar [dd] ^{\phi^{*}Q' (C' \hookrightarrow C)}\\
&& \\
\phi^{*}Q(C') \ar [rr] _{\phi^{*} \theta_{C'}} & & \phi^{*}Q' (C') \\
}
\end{equation}
can be easily checked by diagram chasing.
It is obvious that $\phi^{*}$ preserves identities and products.

In order to give the definition of the direct image part
$\phi_{\ast}:\Set^{\Cboldop} \to \Set^{\Vboldop}$,
we define a functor $1_{\downarrow V} \in \Set^{\Vboldop}$,
the subobject of the terminal object $1$ of $\Set^{\Vboldop}$ 
which is not empty only for $V' \subseteq V$:
\begin{equation}
1_{\downarrow V}(V') := 
\begin{cases}
\{ \pt \} & (V' \subseteq V) \\
\emptyset & (V' \not\subseteq V) \\
\end{cases},
\label{eq:1downarrowV}
\end{equation}
where $\{ \pt \} $ is the one point set.
Note that, for each $V' \hookrightarrow V$,
we can define a morphism $1_{\downarrow V'} \xrightarrow{1_{\downarrow (V' \hookrightarrow V)}} 1_{\downarrow V}$
by the natural inclusion:
\begin{equation}
1_{\downarrow (V' \hookrightarrow V)} := 
\xymatrix{
1_{\downarrow V'} \ar @{^{(}->} [r] & 1_{\downarrow V}
}.
\end{equation}
Thus, we can regard (\ref{eq:1downarrowV}) as 
a definition of the bifunctor $1_{\downarrow}:\Vbold \times \Vboldop \to \Set$.

By means of the bifunctor $1_{\downarrow}$,
we define $\phi_{*}$ as follows:
For each $P \in \Set^{\Cboldop}$ and $V \in \Vbold$,
$\phi_{*}P(V) $ is given by
\begin{equation}
\phi_{*}P(V) 
:=
\Hom_{\Cboldhat}(\phi^{*}(1_{\downarrow V}), P) \nonumber\\
=
\Hom_{\Cboldhat}((1_{\downarrow V})_{\phi}, P),
\label{eq:phi*P(V)}
\end{equation}
and, for each $V' \hookrightarrow V$ and $\alpha \in \phi_{*}P(V)$,
$\phi_{*}P(V' \hookrightarrow V)(\alpha) \equiv \alpha|_{V'} \in \phi_{*}P(V')$
is defined as a morphism which makes the diagram
\begin{equation}
\xymatrix{
\phi^{*}(1_{\downarrow V}) \ar [rr] ^{\alpha}  && P \\
&& \\
\phi^{*}(1_{\downarrow V' })  \ar  [uu] ^{\phi^{*}(1_{\downarrow (V' \hookrightarrow V) })} \ar [rruu] _{\phi_{*}P(V' \hookrightarrow V)(\alpha)}&& \\
}
\label{eq:phi_{*}Pobjmor}
\end{equation}
commute.

For each morphism $\xymatrix{P \ar [r] ^{\theta} & P'} \in \Mor(\Set^{\Cboldop})$,
the morphism $\xymatrix{\phi_{*}P \ar [r] ^{\phi_{*}\theta} & \phi_{*}P'}$
is defined by, for every $V \in \Vbold$, 
the map $\xymatrix{\phi_{*}P(V) \ar [r] ^{(\phi_{*}\theta)_{V}} & \phi_{*}P'(V)}$
which assigns the morphism 
$(\phi_{*}\theta)_{V}(\alpha)$ making the diagram
\begin{equation}
\xymatrix{
\phi_{*}(1_{\downarrow V}) \ar [rr] ^{\alpha} \ar [rrdd] _{(\phi_{*}\theta)_{V} (\alpha)} && P \ar [dd] ^{\theta}\\
&& \\
&& P' \\
}
\end{equation}  
commute to every $\alpha \in \phi_{*}P(V)$.
Note that the naturality condition for $\phi_{*}\theta$,
namely, the commutativity of the diagram
\begin{equation}
\xymatrix{
\phi_{*}P(V) \ar [rr] ^{(\phi_{*}\theta)_{V}} \ar [dd] _{\phi_{*} P(V' \hookrightarrow V)} && \phi_{*}P'(V) \ar [dd] ^{\phi_{*} P'(V' \hookrightarrow V)}\\
&& \\
\phi_{*}P(V') \ar [rr] _{(\phi_{*}\theta)_{V'}} && \phi_{*}P' (V')  \, ,\\
}
\end{equation}
is equivalent to that 
of the lower triangle on the following diagram:
\begin{equation}
\xymatrix{
\phi^{*}(1_{\downarrow V}) \ar [rrr] ^{\alpha} &&& P\, \ar [dd] ^{\theta} \\
&&& \\
\phi^{*}(1_{\downarrow V'}) \ar [uu] ^{\phi^{*}(1_{\downarrow(V' \hookrightarrow V)})} 
\ar [rrruu] _{\phi_{*}P(V' \hookrightarrow V)(\alpha)} 
\ar [rrr] _-{\phi_{*}P(V' \hookrightarrow V)(\theta \circ \alpha)}&&& P'\,. \\
}
\end{equation} 
Indeed, it commutes since
the outer square and the upper triangle commute
because of (\ref{eq:phi_{*}Pobjmor}).
Thus, the functor  $\phi_{*}$ is well-defined.

\section{Topologies on Quantum Topos Induced by Quantization Functors}
\label{sec:Topologies on Quantum}

It is well-known that every geometric morphism induces  
a Lawvere-Tierney topology on its target topos
and that any Lawvere-Tierney topology on a presheaf topos is
equivalent to a Grothendieck topology on the context category of the presheaf topos~\cite{MM92}.
In the present section,
we derive Lawvere-Tierney topologies on $\Set^{\Vboldop}$
and corresponding Grothendieck topologies on $\Vbold$
induced by the geometric morphisms $\phi$ given in the previous section.
Definitions of the Lawvere-Tierney topology, the Grothendieck topology
and the related closure operator are given in Appendix \ref{sec:Lawvere-Tierney Topology, Grothendieck}.

\subsection{Closure operator induced by geometric morphism}
\label{subsec:Closure operator induced}

Following the method,
we start by constructing a closure operator
induced by $\phi$.
As is explained in Appendix \ref{sec:Lawvere-Tierney Topology, Grothendieck},
a closure operator $\overline{\cdot}$ is a map on $\Sub_{\Vboldhat}(Q)$
($Q \in \Set^{\Vboldop}$)
satisfying  conditions (\ref{eq:closurecondition}).
A geometric morphism $\phi$ defines a closure operator as follows:
For any $Q \in \Set^{\Vboldop}$ and $\xymatrix{S \ar @{>->} [r] ^{\iota} & Q}$,
the closure $\xymatrix{\overline{S} \ar @{>->} [r] ^{\overline{\iota}} & Q}$  
is the pullback of the monomorphism 
$\xymatrix{ \phi_{\ast}\phi^{\ast} S \ar @{>->} [r] ^{\phi_{\ast}\phi^{\ast} \iota} & \phi_{\ast}\phi^{\ast} Q}$,
\begin{equation}
\xymatrix{
\overline{S} \ar [rr] ^-{\xi} \ar @{>->} [dd] _{\overline{\iota}}  && \phi_{\ast}\phi^{\ast} S  \ar @{>->} [dd]  ^{\phi_{\ast}\phi^{\ast} \iota} \;\\
&& \\
Q    \ar [rr] _-{\eta_{Q}}      &&  \phi_{\ast}\phi^{\ast} Q   \,, \\
} 
\label{eq:closure}
\end{equation}
along the morphism $\eta_{Q}$ given by (\ref{eq:eta}).

Let us mention a few points.
That the morphism $\phi_{*}\phi^{*} \iota$ is monic 
comes from the fact that $\phi_{*}$ and $\phi^{*}$
are left exact;
such functors between topoi always preserve a monomorphism
because the latter is an equalizer in a topos.
In the current case, of course, we can check that by direct calculation
from the definitions of $\phi_{*}$ and $\phi^{*}$ given in the previous section.
Also, it is not difficult to check that definition (\ref{eq:closure})
indeed satisfies (\ref{eq:closurecondition}).
In our case, of course, it is easier to use
the explicit expression of the closure which we give below.
In order to calculate the closure in (\ref{eq:closure}), 
it is enough to assume that the monomorphism $\xymatrix{S \ar @{>->} [r] ^{\iota} & Q}$
is an inclusion $\xymatrix{S \ar @{^{(}->} [r] ^{\iota} & Q}$, that is,
$S(V) \subseteq Q(V)$ for all $V \in \Vbold$.
This is because, for any $T \in \Sub(Q)$,
there exists an inclusion $S \in \Sub(Q)$ such that $S \simeq T$,
and for them, $\phi_{*} \phi^{*} S \simeq \phi_{*} \phi^{*} T$,
hence $\overline{S} \simeq \overline{T}$.
 
We can obtain $\overline{S}$ by solving the pullback diagram for every $V \in \Vbold$.
That is, for every $V$,
the diagram
\begin{equation}
\xymatrix{
\overline{S}(V) \ar [rr] ^-{\xi_{V}} \ar @{>->} [dd] _{\overline{\iota}_{V}}  && \Hom_{\Cboldhat} ( (1_{\downarrow V}) _{\phi},S_{\phi})  \ar @{>->} [dd]  ^{\Hom_{\Cboldhat} ( (1_{\downarrow V}) _{\phi},\iota_{\phi}) } \\
&& \\
Q(V)    \ar [rr] ^-{\eta_{Q(V)}}      &&  \Hom_{\Cboldhat} ( (1_{\downarrow V}) _{\phi},Q_{\phi})  \\
}
\label{eq:closurecomponent}
\end{equation}
is a set theoretic pullback,
and the collection of all solutions $\overline{S}(V)$ 
gives  $\overline{S}$.

We can readily solve (\ref{eq:closurecomponent}) as
\begin{equation}
\overline{S}(V) 
:= 
\{ q \in Q(V) \;|\; \forall C \in \Cbold 
\;(\phi(C) \subseteq V \Rightarrow q|_{\phi(C)} \in S(\phi(C))) \}.
\label{eq:closuresolution}
\end{equation}
This solution indeed gives a subobject of $Q$.
Let $q \in \overline{S}(V)$. Then, $q|_{\phi(C)} \in S(\phi(C))$ 
whenever $\phi(C) \subseteq V $ by the definition. 
For any $V' \hookrightarrow V$, consider $q|_{V'} \equiv Q(V' \hookrightarrow V)(q) \in Q(V')$.
For any $C'$ such that $\phi(C') \subseteq V'$, we have 
$q|_{V'}|_{\phi(C')} = q|_{\phi(C')} \in S(\phi(C'))$ because $\phi(C') \subseteq V' \subseteq V$.
Therefore, $q|_{V'} \in \overline{S}(V')$. 
Thus, (\ref{eq:closuresolution}) and 
$\overline{S}(V' \hookrightarrow V) := Q(V' \hookrightarrow V)|_{\overline{S}(V)}$ define
$\overline{S} \in \Sub(Q)$.
The morphism $\xymatrix{\overline{S} \ar @{>->} [r] ^{\overline{\iota}} & Q}$
is defined as an inclusion.

The natural transformation $\xi$ is defined by
\begin{equation}
\xi_{V}(q)_{C}(\pt) := q|_{\phi(C)} \in S_{\phi}(C),
\label{eq:xi}
\end{equation}
for each $V \in \Vbold$, $C \in \Cbold$ such that $\phi(C) \subset V$,
and $q \in S(V)$.
It is easy to show that 
(\ref{eq:xi})
indeed defines a morphism in $\Set^{\Vboldop}$; 
that is, $\xi$ satisfies the required naturality conditions
corresponding to (\ref{eq:taucondition1}) and (\ref{eq:taucondition2}) given in appendix \ref{subsec:Adjunction}.

Under the above-mentioned definitions of $\overline{S}$ and $\xi$,
we can show that the diagram (\ref{eq:closure}) is a pullback.
First, note that the commutativity of (\ref{eq:closure}) is equivalent to
that of the diagram
\begin{equation}
\xymatrix{
(1_{\downarrow V})_{\phi} \ar [rr] ^-{(\eta_{Q})_{V}(q)} \ar [rdd] _{\xi_{V}(q)} & & Q_{\phi} \\
&& \\
 & \;\; S_{\phi} \ar [ruu] _{\iota_{\phi}} \;.& \\
}
\end{equation}
This diagram in fact commutes because,
whenever $\phi(C) \subseteq V$,
\begin{equation}
(\iota_{\phi})_{C} \circ \xi_{V}(q)_{C} (\pt) = q|_{\phi(C)}= (\eta_{Q})_{V}(q)_{C}(\pt).
\end{equation}

Next, Let $\xymatrix{T \ar [r] ^{\alpha} & Q}$ and
$\xymatrix{T \ar [r] ^{\beta}& \phi_{*}\phi^{*}}S$ make the outer square of the diagram
\begin{equation}
\xymatrix{
T \ar [rrrd] ^{\beta} \ar [rddd] _{\alpha} \ar @{.>} [rd] ^{\gamma} & & \\
& \overline{S} \ar [rr] ^-{\xi} \ar @{>->} [dd] _{\overline{\iota}}  && \phi_{\ast}\phi^{\ast} S  \ar @{>->} [dd]  ^{\phi_{\ast}\phi^{\ast} \iota} \\
&& \\
& Q    \ar [rr] ^{\eta_{Q}}      &&  \phi_{\ast}\phi^{\ast} Q    \\
}
\end{equation}
commute;
that is, suppose that, for any $x\in T(V)$, the diagram
\begin{equation}
\xymatrix{
(1_{\downarrow V})_{\phi} \ar [rr] ^-{(\eta_{Q})_{V}(\alpha_{V}(x))} \ar [rdd] _{\beta_{V}(x)} & & Q_{\phi} \\
&& \\
 & S_{\phi} \ar [ruu] _{\iota_{\phi}}& \\
}
\end{equation}
commutes.
This implies that, for any $C \in \Cbold$ such that $\phi(C) \subseteq V$,
\begin{equation}
(\eta_{Q})_{V}(\alpha_{V}(x))_{C}(\pt)
=  \alpha_{V}(x)|_{\phi(C)} = \beta_{V}(x)_{C}(\pt) \in S_{\phi}(C),
\end{equation}
and hence, $\alpha_{V}(x) \in \overline{S}(V)$ from definition (\ref{eq:closuresolution}) of $\overline{S}$. 

For each $V \in \Vbold$, we define a function $\gamma_{V}:T(V) \to \overline{S}(V)$ by
\begin{equation}
\gamma_{V}(x):= \alpha_{V}(x) .
\end{equation}
Then, it is clear that $\alpha_{V} = \overline{\iota}_{V} \circ \gamma_{V}$.
We can show that $\gamma_{V}$'s are natural in $V$.
Therefore, they give a natural transformation 
$\xymatrix{T \ar [r] ^{\gamma} & \overline{S}}$ which satisfies 
\begin{equation}
\overline{\iota} \circ \gamma = \alpha.
\label{eq:gamma1}
\end{equation}
Furthermore, we have
\begin{eqnarray}
\xi_{V}(\gamma_{V}(x))_{C}(\pt)
& = &
\gamma_{V}(x)|_{\phi(C)} \nonumber\\
& = &
\alpha_{V}(x)|_{\phi(C)} \nonumber\\
& = &
(\eta_{Q})_{V}(\alpha_{V}(x))_{C}(\pt) \nonumber\\
& = &
\beta_{V}(x)_{C}(\pt) , 
\end{eqnarray} 
hence,
\begin{equation}
\xi \circ \gamma = \beta.
\label{eq:gamma2}
\end{equation}
Morphisms $\gamma$ satisfying (\ref{eq:gamma1}) and (\ref{eq:gamma2}) are unique
because $\overline{\iota}$ is monic.
It is thus shown that (\ref{eq:closuresolution}) is indeed the solution
of pullback diagram (\ref{eq:closure}).

\subsection{Grothendieck topology and Lawvere-Tierney topology}
\label{subsec:Grothendieck topology and}

As is explained in Appendix \ref{sec:Lawvere-Tierney Topology, Grothendieck},
the Grothendieck topology $J$ is defined by the closure 
of terminal object $1$ as a subobject of the subobject classifier $\Omega$:
\begin{equation}
J = \overline{1}.
\label{eq:Grothedieck}
\end{equation}
More precisely, it is a solution of pullback diagram (\ref{eq:closure})
for  $\xymatrix{S \ar @{>->} [r] ^{\iota} & Q}=\xymatrix{1 \ar @{>->} [r] ^{\true}  & \Omega}$.
Regarding the terminal object $1$ as
\begin{equation}
1(V) = \{ \tmath_{V} \} \subseteq \Omega (V),
\end{equation}
we obtain an expression of $J(V)$ ($V \in \Vbold$)
from equality (\ref{eq:closuresolution}):
\begin{eqnarray}
J(V) & = & \overline{1}(V) \nonumber\\
& = &
\left\{ 
\omega \in \Omega(V) \;|\; 
\forall C \in \Cbold 
\left( 
\phi(C) \subseteq V  
\Rightarrow 
\omega|_{\phi(C)} \in 1(\phi(C))
\right)
\right\} \nonumber \\
& = &
\left\{ 
\omega \in \Omega(V) \;|\; 
\forall C \in \Cbold 
\left( 
\phi(C) \subseteq V  
\Rightarrow 
\omega|_{\phi(C)} = \tmath_{\phi(C))}
\right)
\right\} \nonumber \\
& = &
\left\{ 
\omega \in \Omega(V) \;|\; 
\forall C \in \Cbold 
\left( 
\phi(C) \subseteq V  
\Rightarrow 
\phi(C) \in \omega|_{\phi(C)}
\right)
\right\}  \nonumber \\
& = &
\left\{ 
\omega \in \Omega(V) \;|\; 
\forall C \in \Cbold 
\left( 
\phi(C) \subseteq V  
\Rightarrow 
\phi(C) \in \omega
\right)
\right\} ,
\label{eq:Grothendieckcomponent} .
\end{eqnarray}
Of course, as a subobject of $\Omega$,
$J(V' \hookrightarrow V) = \Omega (V' \hookrightarrow V)|_{J(V)}$.
Also, the morphism $\xymatrix{J \ar @{>->} [r] ^{\overline{\true}} & \Omega}$ 
is an inclusion.
It is easy to check that  (\ref{eq:Grothendieckcomponent})
indeed satisfies the definition given 
in Appendix \ref{sec:Lawvere-Tierney Topology, Grothendieck}.

The Lawvere-Tierney topology $\xymatrix{\Omega \ar [r] ^{j} & \Omega}$ 
is just the characteristic morphism of  $J$;
that is, $j$ makes diagram (\ref{eq:J-j}) a pullback.
Thus, each component of $j$, 
$\Omega(V) \xrightarrow{j_{V}} \Omega(V)$ ($V \in \Vbold$), 
is given by
\begin{eqnarray}
j_{V}(\omega) & = & 
\left\{ 
V' \in \Sub(V)
\;|\; 
\omega|_{V'} \in J(V')
\right\} \nonumber \\
& = &
\left\{ 
V' \in \Sub(V) \;|\; 
\forall C \in \Cbold 
\left( 
\phi(C) \subseteq V'  
\Rightarrow 
\phi(C) \in \omega
\right)
\right\}.
\label{eq:LawvereTierney}
\end{eqnarray}
It is not difficult to check that $j$ defined by
(\ref{eq:LawvereTierney}) indeed satisfies the conditions of 
Lawvere Tierney topology given by (\ref{eq:LawvereTierneycondition1})-(\ref{eq:LawvereTierneycondition3}).

\subsection{The coarsest topology on the quantum topos}
\label{subsec:The coarsest topology}

It is obvious, from (\ref{eq:Grothendieckcomponent}), 
that larger prequantization topoi induce coarser 
Grothendieck topologies on $\Set^{\Vboldop}$:
That is, for $\Cbold_{1}$, $\Cbold_{2} \in \Cmath$,
and the corresponding topologies $J_{1}$ and $J_{2}$,
we have
\begin{equation}
\Cbold_{1} \subseteq \Cbold_{2}
\implies
J_{2} \subseteq J_{1},
\label{eq:C1C2J2J1}
\end{equation}
where the right hand side means the presheaf inclusion.
Or equivalently, in terms of Lawvere-Tierney topologies $j_{1}$ and $j_{2}$,
we have
\begin{equation}
\Cbold_{1} \subseteq \Cbold_{2}
\implies
j_{2} \preceq  j_{1},
\label{eq:C1C2j2j1}
\end{equation}
where 
$j_{2} \preceq  j_{1}$ means that
$(j_{2})_{V}(\omega) \subseteq (j_{1})_{V}(\omega) $
for all $V \in \Vbold$ and for all $\omega \in \Omega(V)$.

We should note that larger prequantization categories have more information
about the classical system. 
We can say, therefore, 
that coarser topologies bring more information about it
into the quantum topos $\Set^{\Vboldop}$.
One might think that the largest one, $\Cbold_{\Ocal}$, is the most informative, unique prequantization category
which induce the coarsest topology.
However, it is actually redundant;
we can obtain the coarsest topology by means of smaller categories than $\Cbold_{\Ocal}$.
That is, we have the following theorem:
\begin{thm}
Any prequantization category $\Cbold \in \Cmath$ induces the same topology
on the quantum topos $\Set^{\Vboldop}$,
whenever $\Cbold$ includes $\Abold$ as a subcategory.
The induced closure operator, the Grothendieck topology, 
and the Lawvere-Tierney topology
are given by
\begin{equation}
\overline{S} (V) = \{ q \in Q(V) \,|\, q|_{\flat(V)} \in S(\flat(V))\} ,
\label{eq:cl}
\end{equation}
\begin{equation}
J(V) 
=
\{
\omega \in \Omega(V) \,|\, \flat(V) \in \omega \},
\label{eq:Grothen}
\end{equation}
and
\begin{equation}
j_{V}(\omega)
=
\{
V' \in \Sub(V) \,|\, \flat(V') \in \omega \} ,
\label{eq:LT}
\end{equation}
respectively. 
\end{thm}   
{\it Proof.} 
If $\Cbold$ includes $\Abold$,
we can delete the quantifiers $\forall$ from the expressions for 
$\overline{S}$, $J$, and $j$ given by (\ref{eq:closuresolution}),
(\ref{eq:Grothendieckcomponent}), and (\ref{eq:LawvereTierney}).
For example,
let us derive (\ref{eq:cl}) from (\ref{eq:closuresolution}).
Suppose that an element $q$ of $Q(V)$
satisfies $q|_{\flat(V)} \in S(\flat(V))$.
Then, for each $C \in \Cbold$ such that $\phi(C) \subseteq V$,
we have $q|_{\phi(C)} \in S(\phi(C))$,
because $C \subseteq \psi(V)$ from the Galois connection relation.
Therefore $q|_{\phi(C)} = q|_{\flat(V)}|_{\phi(C)} $.
Conversely, suppose that $q \in \overline{S}(V)$ 
where $\overline{S}(V)$ is given by (\ref{eq:closuresolution})
then, especially for $C= \psi(V) \subseteq V$,
which is an object of $\Cbold$ since $\Abold \subseteq \Cbold$,
we have $q|_{\phi(\psi(V))}=q|_{\flat(V)} \in S(\flat(V))$.
Thus, it follows that (\ref{eq:closuresolution}) and (\ref{eq:cl}) 
are the same when $\Abold \subseteq \Cbold$.
Also, (\ref{eq:Grothen}) and (\ref{eq:LT}) can be derived similarly.
\hspace{\fill}$\square$

Consequently, the most informative topology 
in the quantum topos $\Set^{\Vboldop}$ induced by 
quantization is given by the quite simple expressions given by (\ref{eq:cl})-(\ref{eq:LT}),
and $\Abold$ is the least prequantization category which induces them. 
We give the following definition:
\begin{dfn}
We call a topology on a quantum topos $\Set^{\Vboldop}$ given by 
(\ref{eq:cl}), (\ref{eq:Grothen}), and (\ref{eq:LT}) 
a quantization topology.
\end{dfn}

\subsection{Generalized quantization topology}
\label{subsec:Generalized quantization topology}

In the previous subsection, we defined the quantization topology.
This notion, however, can be extended in such a way that does not need 
prequantization categories and topoi.

To do so, we note that we defined the quantization topology 
by means of only the endofunctor $\flat:\Vbold \to \Vbold$
without referring to prequantization categories. 
This means that any quantization topology on $\Set^{\Vboldop}$
can be constructed without data about prequantization categories.

As was defined by equation (\ref{eq:sharpflat}),
the endofunctor $\flat$ is given by $\flat = \phi \circ \psi$,
where the pair of the quantization functor $\phi$ and the classicization functor $\psi$
is a Galois connection;
that is, they are an adjunction $\phi \dashv \psi$.
The endofunctor $\flat$, therefore,
defines a comonad $(\flat, \delta, \epsilon)$ on $\Vbold$~\cite{MM92}.
That is, we can give the natural transformations $\delta$ and $\epsilon$
which make the definition diagrams  
\begin{equation}
\xymatrix{
\flat(V) \ar [rr] ^{\delta_{V}} \ar [dd] _{\delta_{V}} && \flat^{2}(V) \ar [dd] ^{\delta_{\flat(V)}} \\
&& \\
\flat^{2}(V) \ar [rr] _{\flat \delta_{V}} && \flat^{3}(V) \\
}
\end{equation}
and 
\begin{equation}
\xymatrix{
             && \flat(V) \ar @{=} [lldd] \ar [dd] ^{\delta_{V}} \ar @{=} [rrdd] && \\
             && \\
\flat(V) && \flat^{2}(V) \ar [ll] ^{\epsilon_{\flat(V)}} \ar [rr] _{\flat \epsilon_{V}}&& \flat(V) \\
}
\end{equation}
of a comonad commute.
In the above diagrams, since $\flat^{2} = \flat$ and hence $\flat^{3} = \flat$,
we can define $\delta$ is an identity,
whereas $\epsilon$ is given by $\flat \leq  I$.

Every quantization Grothendieck (hence, Lawvere-Tierney) topology
is defined by a comonad $\flat := (\flat, \flat^2=\flat, \flat \leq  I)$ on $\Vbold$.
We, therefore, give the following definition:
\begin{dfn}
We call a comonad $\flat$ on $\Vbold$ 
a generalized quantization topology on $\Set^{\Vboldop}$.
\end{dfn} 

As previously explained, each quantization functor
determines a generalized quantization topology $\flat$ uniquely.
Conversely, each generalized quantization topology $\flat$
defines topologies on the quantum topos $\Set^{\Vboldop}$
via definitions (\ref{eq:cl}), (\ref{eq:Grothen}), and (\ref{eq:LT}),
though it is not clear if there is a classical system
and a quantization map which induce the topology given by
arbitrary generalized quantization topos.

For arbitrary $\flat$,
we can always construct two categories, i.e., co-Eilenberg-Moore category $\Vbold_{\flat}$
and co-Kleisli category $ \Vbold^{\flat}$.
According to the prescription for its construction,
the former is the subcategory $\Vbold_{\flat}$ defined by (\ref{eq:Vflat}).
And the Lawvere-Tierney topology on $\Set^{\Vboldop}$
is derived from the geometric morphism from $\Set^{\Vboldop_{\flat}}$
to $\Set^{\Vboldop}$ induced by the inclusion functor $\Vbold_{\flat} \hookrightarrow \Vbold$.
As was noted in Section \ref{subsec:Proper prequantization category},
if $\flat$ is given by a quantization,
the co-Eilenberg-Moore category is isomorphic to the proper prequantization category $\Abold$.

\section{Quantization Sheaf Topos}
\label{sec:Quantization Sheaf Topos}

In the last section,
we showed that any prequantization category $\Cbold$
including the proper prequantization category $\Abold$
induces the same topology,
the quantization topology defined by
(\ref{eq:LT}) (or equivalently, (\ref{eq:Grothen}) or (\ref{eq:cl})),
on the quantum topos $\Set^{\Vboldop}$.

Throughout this section, we suppose that we are given 
a quantization topology defined by  (\ref{eq:LT}).
Our purpose in this section is to explain how the quantization topology
induces sheaves and a topos of sheaves
in the quantum topos $\Set^{\Vboldop}$.
It should be noted that there exists a well-established,
sophisticated (and rather abstract) general theory to construct
a sheaf topos from a given topology~\cite{MM92,Bor94c}.
In this section, however, 
we intend to give an elementary and self-contained construction
by use of the simple expression of the quantization topology.
That would show clearly the structure of the sheaf topos,
and would be useful for application.

\subsection{Endofunctor $\flat^{*}$}
\label{subsec:Endofunctor}

The endofunctor $\flat:\Vbold \to \Vbold$ induces
the geometric morphism 
$\flat:\Set^{\Vboldop} \to \Set^{\Vboldop}$.
In this section, the inverse image part 
$\flat^{*}:\Set^{\Vboldop}\to \Set^{\Vboldop}$
plays a crucial role.
So, for convenience of reference, we give the definition
in the following:
For each presheaf $Q \in \Set^{\Vboldop}$,
the presheaf $\flat^{*}Q$ is defined by
\begin{equation}
\flat^{*} Q \equiv Q_{\flat} := Q \circ \flat^{\rm op},
\label{eq:sheafification}
\end{equation}
namely, for each $V \in \Vbold$,
\begin{equation}
\flat^{*} Q (V) \equiv Q_{\flat}(V) := Q(\flat(V)),
\label{eq:sheafificationObj1}
\end{equation}
and, for each $V' \hookrightarrow V$ and $q^{\flat} \in Q_{\flat}(V) = Q(\flat(V))$,
\begin{eqnarray}
\flat^{*} Q (V' \hookrightarrow V) (q^{\flat}) 
& \equiv & 
Q_{\flat}  (V' \hookrightarrow V)  (q^{\flat})  \equiv q^{\flat}||_{V'} \nonumber\\
& := &
Q(\flat(V') \hookrightarrow \flat(V))(q^{\flat}) \equiv q^{\flat}|_{\flat(V')} .
\label{eq:sheafificationObj2}
\end{eqnarray}
Also,
for every $\xymatrix{Q \ar [r] ^{\theta} & Q'} \in \Mor (\Set^{\Vboldop})$,
the morphism $\xymatrix{\flat^{*}Q \ar [r] ^{\flat^{*}\theta} & \flat^{*} Q'} 
\equiv \xymatrix{Q_{\flat} \ar [r] ^{\theta_{\flat}} & Q'_{\flat} }\in \Mor (\Set^{\Vboldop})$
is defined by, for each $V \in \Vbold$ and $q^{\flat} \in \flat^{*}Q(V)$,
\begin{eqnarray}
\flat^{*}\theta_{V} (q^{\flat})
\equiv (\theta_{\flat})_{V}(q^{\flat})
:=
\theta_{\flat(V)} (q^{\flat}) \in Q(\flat(V))= \flat^{*}Q(V).
\label{eq:sheafificationMor}
\end{eqnarray}
It should be noted that $\flat^{*}$ is left exact as the inverse image part 
of the geometric morphism $\flat$.
This fact is used in \S \ref{subsec:Topos conditions}.

In this section, moreover, we need the natural transformation 
$\xymatrix{I \ar [r] ^{\zeta} & \flat^{*}}$
whose $Q$-component $\xymatrix{Q \ar [r] ^{\zeta_{Q}} & \flat^{*}Q}$
is given by
\begin{equation}
(\zeta_{Q})_{V}(q) := Q(\flat(V) \hookrightarrow V)(q) \equiv q|_{\flat(V)} \in \flat^{*}Q(V).
\label{eq:zetadef}
\end{equation}
It is easy to check that $\zeta_{Q}$ is indeed a morphism in $\Set^{\Vboldop}$;
that is, it makes the diagram
\begin{equation}
\xymatrix{
Q(V) \ar [rr] ^{(\zeta_{Q})_{V}} \ar [dd] _{Q(V' \hookrightarrow V)} && 
\flat^{*} Q(V) \ar [dd] ^{\flat^{*}Q(V' \hookrightarrow V)} \\
&& \\
Q(V') \ar [rr] _{(\zeta_{Q})_{V'}} && \flat^{*}Q(V') \\
}
\end{equation}
commute.
Also, $\zeta$ is indeed a natural transformation 
from the identity functor $I$ to $\flat^{*}$;
that is, for every $\xymatrix{Q \ar [r]^{\theta} & Q'}$, the diagram
\begin{equation}
\xymatrix{
Q \ar [rr] ^{\zeta_{Q}} \ar [dd] _{\theta} && \flat^{*}Q \ar [dd] ^{\flat^{*} \theta} \\
&& \\
Q' \ar [rr] _{\zeta_{Q'}}  && \flat^{*} Q' \\
}
\label{eq:zetanaturality}
\end{equation}
commutes.

As a counterpart of the right hand equality on (\ref{eq:sssfff}),
we have the following equality:
\begin{prp}
The endofunctor $\flat^{*}$ is an idempotent:
\begin{equation}
\flat^{*}\flat^{*} = \flat^{*}.
\end{equation}
More precisely, the diagram
\begin{equation}
\xymatrix{
&& \flat^{*}\flat^{*} \ar @{=} [dd] \\
\flat^{*} \ar [rru] ^{\flat^{*} \zeta} \ar @{=} [rrd]  & &   \\
&& \flat^{*} \\
}
\label{eq:flat*flat*=flat*}
\end{equation}
commutes;
$\flat^{*} \zeta$ is the identity natural transformation
from the functor $\flat^{*}$ to itself.
\end{prp}
{\it Proof.}
For every $Q \in \Set^{\Vboldop}$,
\begin{equation}
\flat^{*}\flat^{*}Q
 = 
\flat^{*}Q_{\flat} 
 = 
Q_{\flat\flat} 
 = 
Q_{\flat} 
 = 
\flat^{*} Q.
\end{equation}
Furthermore,
for each $q^{\flat} \in \flat^{*} Q(V)$,
\begin{eqnarray}
(\flat^{*}\zeta_{Q})_{V}(q^{\flat})
& = &
(\zeta_{Q})_{\flat(V)}(q^{\flat}) \nonumber\\
& = &
Q(\flat(V) \hookrightarrow \flat(V)) (q^{\flat}) \nonumber\\
& = &
q^{\flat}.
\end{eqnarray}
Thus, the proposition is proved. \hspace{\fill}$\square$

\subsection{Sheaf conditions}
\label{subsec:Sheaf conditions}

First we recall the definition of a sheaf ~\cite{MM92}:
\begin{dfn}
\label{dfn:quantizationsheaf}
A presheaf $R \in \Set^{\Vboldop}$ is called a sheaf 
for the quantization topology $j$ (or, j-sheaf) if and only if,
for every presheaf $Q \in \Set^{\Vboldop}$, 
every its subobject $S$ of $Q$ which is dense in $Q$,
and every morphism $\xymatrix{S \ar [r] ^{\alpha} & R}$,
there exists a unique morphism $\xymatrix{Q \ar [r] ^{\alphabreve} & R}$ 
which makes the diagram
\begin{equation}
\xymatrix{
S \ar [rr] ^{\alpha} \ar @{>->} [dd] _{{\rm dense}} && R \\
&& \\
Q \ar  [rruu] _{ \alphabreve}&& \\
}
\label{eq:dfn:sheaf}
\end{equation}
commute;
namely, there exists an isomorphism
\begin{equation}
\Hom_{\Vboldhat}(S,R) \simeq \Hom_{\Vboldhat}(Q,R).
\end{equation}
\end{dfn}
Here, $S \in \Sub(Q)$ is called dense in $Q$ when
\begin{equation}
\overline{S} = Q.
\label{eq:dense}
\end{equation} 
\begin{dfn}
We call a $j$-sheaf a {\it quantization sheaf}
if the topology $j$ is a quantization topology.
\end{dfn}
%
Our main purpose in this subsection is to prove the following statement:
\begin{thm}
\label{thm:sheafcondition}
A presheaf $R \in \Set^{\Vboldop}$ is a quantization sheaf
if and only if 
the morphism $\xymatrix{R \ar [r] ^{\zeta_{R}} & \flat^{*}R}$
is isomorphic.
\end{thm}
Definition (\ref{eq:dfn:sheaf}) of a sheaf implies that,
if $R$ is a sheaf and a presheaf $F$ is isomorphic to $R$,
then as well $F$ is a sheaf.
Therefore, in order to prove the above theorem,
it is enough to verify the following propositions:
\begin{prp}
\label{prp:sheafsufficientcondition}
For every presheaf $F$, $\flat^{*}F$ is a quantization sheaf.
\end{prp}
\begin{prp}
\label{prp:sheafnecessarycondition}
For every quantization sheaf $R$, $R \xrightarrow{\zeta_{R}} \flat^{*}R$ is an isomorphism.
\end{prp}

{\it Proof of Proposition \ref{prp:sheafsufficientcondition}.}
Let $Q$ be an arbitrary presheaf and $S \hookrightarrow Q$ be
a dense subobject of $Q$, namely, $\overline{S} = Q$.
Because of definition (\ref{eq:cl}) of $\overline{S}$,
$S$ is dense in $Q$ if and only if, for all $V \in \Vbold$,
\begin{eqnarray}
q \in Q(V)
& \iff &
q \in \overline{S}(V) \nonumber\\
& \iff &
q|_{\flat(V)} \in S(\flat(V)) = S_{\flat}(V) .
\label{eq:densecondition}
\end{eqnarray}
In particular,
note that, 
if $S$ is dense in $Q$,
we have
\begin{equation}
Q(\flat(V)) = S(\flat(V)) .
\label{eq:Qflat=Sflat}
\end{equation}

Now, let us assume that we have a morphism 
$\xymatrix{S \ar [r] ^{\alpha} & \flat^{*}F}$.
Then, we can construct a unique morphism 
$\xymatrix{Q \ar [r] ^{\alphabreve} & \flat^{*}F}$
which makes the diagram
\begin{equation}
\xymatrix{
S \ar [rr] ^{\alpha} \ar @{>->} [dd] && \flat^{*}F \\
&& \\
Q \ar  [rruu] _{\alphabreve} && \\
}
\label{eq:sheafdiagram2}
\end{equation}
commute.
This is defined by
\begin{equation}
\alphabreve_{V}(q) := \alpha_{V}(q|_{\flat(V)}) \in \flat^{*}F(V) 
\end{equation}
for each $V \in \Vbold$ and $q \in Q(V)$.
It is obvious that $\alphabreve(V)$ defines a natural transformation 
$\alphabreve$.

To see the uniqueness of $\alphabreve$,
suppose that there exists another morphism $\alphabreve'$ which makes the diagram (\ref{eq:sheafdiagram2}) commute.
Then, both of $\alphabreve$ and $\alphabreve'$ 
have to make the diagram 
\begin{equation}
\xymatrix{
Q(V)  \ar [rr] ^{\alphabreve_{V} , \; \alphabreve'_{V}} \ar [ddd] _{Q(\flat(V) \hookrightarrow V)} &     &  \flat^{*}F(V) \ar @{=} [ddd] \\
         & S(V) \ar @{>->} [lu] \ar [ru] _{\alpha_{V}} \ar @{=} [d]  &    \\
         & S(\flat(V)) \ar @{=} [ld] \ar [rd] ^{\alpha_{\flat(V)}}   &    \\
Q(\flat(V)) \ar [rr] _{\alphabreve_{\flat(V)} , \; \alphabreve'_{\flat(V)}}  &                  &           \flat^{*}F(\flat(V))       \\
}
\label{eq:naturalityof alphas}
\end{equation}
commute, 
since it is their naturality conditions
applied to $\flat(V) \hookrightarrow V$.
Here, in diagram (\ref{eq:naturalityof alphas}),
(\ref{eq:flat*flat*=flat*}) and (\ref{eq:Qflat=Sflat}) are reflected.
Note that, from the bottom triangle,
we have
\begin{equation}
\alphabreve_{\flat(V)} = \alpha_{\flat(V)} = \alphabreve'_{\flat(V)} .
\end{equation}
Therefore, the commutativity of the outer square ensures that $\alphabreve_{V} = \alphabreve'_{V}$.
\hspace{\fill}$\square$

Before proving Proposition \ref{prp:sheafnecessarycondition},
we need some preparations.
For each $\omega \in J(V)$,
let us define a subobject $1_{\omega}$ of $1_{\downarrow V}$ by
\begin{equation}
1_{\omega}(V') 
:= 
\begin{cases}
1_{\downarrow V}(V') & (V' \in \omega) \\
\emptyset & (V' \not\in \omega) \\
\end{cases}
.
\label{eq:1omega}
\end{equation}
Note that each $1_{\omega}$ is dense in $1_{\downarrow V}$.
In fact, for every $V' \subseteq V$, it follows that
\begin{eqnarray}
\overline{1_{\omega}}(V') \ne \emptyset
& \iff &
1_{\omega}(\flat(V')) \ne \emptyset \nonumber\\
& \iff &
\flat(V') \in \omega ,
\end{eqnarray}
whereas the last condition is always true if $V' \subseteq V$.
This is 
because $\omega \in J(V)$ if and only if $\flat(V) \in \omega$,
and if $V' \subseteq V$ then $\flat(V') \subseteq \flat(V)$.
Thus, for all $V' \subseteq V$, it holds that 
$\overline{1_{\omega}}(V') \ne \emptyset$, hence,
\begin{equation}
\overline{1_{\omega}} = 1_{\downarrow V} .
\label{eq:1omegadense}
\end{equation}

Now, if a presheaf $R$ is also a quantization sheaf,
then it needs to satisfy
the following condition,
the definition of a sheaf for the quantization Grothendieck topology $J$
(or, $J$-sheaf).
\begin{dfn}
\label{dfn:J-sheaf}
A presheaf $R$ is called a $J$-sheaf (or a sheaf for the Grothendieck topology $J$)
if it satisfies the following condition:
For every $V \in \Vbold$, every $\omega \in J(V)$,
and every morphism $\xymatrix{1_{\omega} \ar [r] ^{\alpha} & R}$,
there exists a unique morphism 
$\xymatrix{1_{\downarrow V} \ar [r] ^{\alphabreve} & R}$ 
which makes the diagram
\begin{equation}
\xymatrix{
1_{\omega} \ar [rr] ^{\alpha} \ar @{>->} [dd] && R \\
&& \\
1_{\downarrow V} \ar  _{\alphabreve} [rruu] & \\
}
\label{eq:dfn:J-sheaf}
\end{equation}
commute;
that is, there exists an isomorphism
\begin{equation}
\Hom_{\Vboldhat}(1_{\omega}, R) \simeq \Hom_{\Vboldhat}(1_{\downarrow V}, R) .
\end{equation}
\end{dfn}
(Note that this definition of a $J$-sheaf is a paraphrase of
the definition given in terms of 
a matching family and an amalgamation~\cite{MM92}.)

{\it Proof of Proposition \ref{prp:sheafnecessarycondition}.}
It is obvious that each $\xymatrix{1_{\downarrow V} \ar [r] ^{\alphabreve} & R}$
is determined by $\alphabreve_{V}(\pt):= r$ for each $r \in R(V)$.
Therefore we have,
\begin{equation}
\Hom_{\Vboldhat}(1_{\downarrow V}, R) \simeq R(V)
\end{equation}
On the other hand, 
the condition for $J$-sheaf need to be true especially for $\omega= \downarrow \flat(V)$,
and also in this case, since $\xymatrix{1_{\downarrow \flat(V)} \ar [r] ^{\alpha} & R}$
is determined by $\alpha_{\flat(V)}(\pt) = r^{\flat} \in R(\flat(V)) $, 
\begin{equation}
\Hom_{\Vboldhat}(1_{\downarrow \flat(V)}, R) \simeq R(\flat(V)) .
\end{equation}
Thus, it follows that
\begin{equation}
R(V) \simeq R(\flat(V)) =  \flat^{*} R(V) .
\label{eq:J-sheafisomorphism}
\end{equation}
Here, the map $\alphabreve \mapsto \alpha$, namely, 
$r \mapsto r^{\flat}$ which gives isomorphism 
(\ref{eq:J-sheafisomorphism}) is $(\zeta_{R})_{V}$.
This is verified from the naturality of diagram (\ref{eq:dfn:J-sheaf}) 
with $\omega = \downarrow \flat(V)$:
\begin{equation}
\xymatrix{
1_{\downarrow V}(V)  \ar [rr] ^{\alphabreve_{V}} \ar @{=} [ddd]  &     &  R(V)\ar [ddd] _{R(\flat(V) \hookrightarrow V) } \ar @{=} [r]
  & R(V) \ar [ddd] ^{(\zeta_{R})_{V}}\\
 & 1_{\downarrow \flat(V)}(V) \ar @{>->} [lu] \ar [ru] _{\alpha_{V}} \ar  [d] _{1_{\downarrow \flat(V)}(\flat(V)\hookrightarrow V)} &   &  \\
 & 1_{\downarrow \flat(V)}(\flat(V)) \ar @{=} [ld] \ar [rd] ^{\alpha_{\flat(V)}}   &    \\
1_{\downarrow V}(\flat(V)) \ar [rr] _{\alphabreve_{\flat(V)} }  &                  &           R(\flat(V))  \ar @{=} [r]
&  \, \flat^{*}R(V) \,,  \\
}
\end{equation} 
where the commutative square added at right hand side 
is the definition of $\zeta$ given by equation (\ref{eq:zetadef}).

Now, chasing this diagram ,
we have that
\begin{eqnarray}
r^{\flat} 
& =  &
\alpha_{\flat(V)}(\pt) \nonumber\\
& = &
\alphabreve_{\flat(V)}(\pt) \nonumber\\
& = &
R(\flat(V) \hookrightarrow V)\circ \alphabreve_{V}(\pt) \nonumber\\
& = &
R(\flat(V) \hookrightarrow V)(r) \nonumber\\
& = &
(\zeta_{R})_{V}(r) .
\end{eqnarray}
Thus, Proposition \ref{prp:sheafnecessarycondition} was proved.
\hspace{\fill}$\square$

It should be noted that,
as was shown previously,
any $j$-sheaf $R^{j}$
has to satisfy the definition of a $J$-sheaf;
namely, a $j$-sheaf is always a $J$-sheaf.
Conversely, we proved that any $J$-sheaf $R^{J}$ makes
$\xymatrix{R^{J} \ar [r] ^{\zeta_{R^{J}}}_{\sim} & \flat R^{J}}$ an natural isomorphism;
a $J$-sheaf is always a $j$-sheaf.
Thus, we are lead to the following well-known fact, which is actually 
true for any Lawvere-Tierney topology $j$ and 
the corresponding Grothendieck topology $J$ on a presheaf topos~\cite{MM92}:
\begin{thm}
The category $\Sh_{j} \Vboldhat $ 
is identical to the category $\Sh_{J} \Vboldhat$.
\end{thm}
Here, $\Sh_{j} \Vboldhat $ (resp. $\Sh_{J} \Vboldhat$) is a full subcategory of $\Set^{\Vboldop}$
whose objects are all $j$-sheaves (resp. all $J$-sheaves).

As was shown above,
the functor $\flat^{*}:\Set^{\Vboldop} \to \Set^{\Vboldop}$
maps every presheaf $Q$ to a quantization sheaf $\flat^{*} Q$.
It is, therefore, a functor from $\Set^{\Vboldop}$ to
the category $\Sh_{j}\Vboldhat$ of sheaves.
Furthermore, 
we can prove the following theorem:
\begin{thm}
The functor $\flat^{*}:\Set^{\Vboldop} \to \Sh_{j}\Vboldhat$ 
is an associated sheaf functor (a sheafification functor).
Namely, it is left adjoint to the inclusion functor
$i:\Sh_{j} \Vboldhat \to \Set^{\Vboldop}$.
\end{thm}
{\it Proof.}
The unit of the adjunction is the natural transformation $\zeta$.
Namely,
we can show that for every morphism $\xymatrix{Q \ar [r] ^{\alpha} & R}$
there exists a unique morphism $\xymatrix{Q \ar [r] ^{\breve{\alpha}} & R}$
making the diagram
\begin{equation}
\xymatrix{
Q \ar [rr] ^{\alpha} \ar [rrdd] _{\zeta_{Q}} && R \\
&& \\
&& \flat^{*} Q \ar [uu] _{\breve{\alpha}} \\
}
\label{eq:sheafificationdef2}
\end{equation}
commute.

The existence directly comes from
diagram (\ref{eq:zetanaturality}),
the naturality condition of $\zeta$,
with $Q' = R$.
Since $R$ is a sheaf, and hence, $\zeta_{R}$ is 
an isomorphism,
we can obtain $\breve{\alpha}$ as
\begin{equation}
\breve{\alpha} := \zeta_{R}^{-1} \circ \flat^{*}\alpha.
\end{equation}

In order to prove the uniqueness,
let us assume that 
there exists another morphism $\xymatrix{\flat^{*}Q \ar [r] ^{\breve{\alpha}'} & R}$ 
which satisfies commutative diagram (\ref{eq:sheafificationdef2}).
What we have to do is to show that the diagram 
\begin{equation}
\xymatrix{
 && R \ar [dd] ^{\zeta_{R}} _{\wr}\\
&& \\
\flat^{*}Q \ar [uurr] ^{\breve{\alpha}'} \ar [rr] _{\flat^{*} \alpha} && \flat^{*}R \\ 
}
\label{eq:sheafification:uniqueness}
\end{equation} 
commutes.
To do so, below we draw the naturality conditions of
$\alpha$, $\flat^{*}\alpha$, $\zeta_{Q}$, $\zeta_{R}$,
and $\breve{\alpha}'$ for $\flat(V) \hookrightarrow V$:
\begin{equation}
\xymatrix{
& Q(V) \ar [rrr] ^{\alpha_{V}} \ar [ld] _{(\zeta_{Q})_{V}} \ar @{.>} [dddd] _{Q(\flat(V) \hookrightarrow V)} &&& R(V) \ar [dddd] ^{R(\flat(V) \hookrightarrow V)} \ar [ld] ^{(\zeta_{R})_{V}} \\
\flat^{*}Q(V) \ar [rrr] _{(\flat^{*}\alpha)_{V}} \ar [rrrru] ^{\breve{\alpha}'_{V}} \ar @{=} [dddd] & && \flat^{*}R(V) \ar @{=} [dddd]  & \\
&&&& \\
&&&& \\
& Q(\flat(V)) \ar @{.>}[rrr] ^{\alpha_{\flat(V)}} 
\ar @{.}@<0.5mm> [ld] \ar @{.}@<-0.5mm> [ld] && & R(\flat(V)) \; \ar @{=} [ld]\\
\flat^{*}Q(\flat(V) ) \ar [rrr] _{(\flat^{*}\alpha)_{\flat(V)}} 
\ar @{.>} [rrrru] ^{\breve{\alpha}'_{\flat(V)}}& && \flat^{*}R(\flat(V) ) \; .& \\
}
\label{eq:box}
\end{equation}
Note that, in the above diagram,
all squares and triangles except for the triangles framed by
$\flat^{*}\alpha$, $\zeta_{R}$, and $\breve{\alpha}'$ are commutative.
Also, the fact that
$\flat^{*}Q(\flat(V) \hookrightarrow V)$, $\flat^{*}R(\flat(V) \hookrightarrow V)$,
$(\zeta_{Q})_{\flat(V)}$, and $(\zeta_{R})_{\flat(V)}$ are identity maps
is reflected.

Now, chasing (\ref{eq:box}) from $q^{\flat} \in \flat^{*}Q(V)$, we have the following equality:
\begin{eqnarray}
(\zeta_{R} \circ \breve{\alpha}')_{V}(q^{\flat})
& = &
(\zeta_{R})_{V}(\breve{\alpha}'_{V}(q^{\flat})) \nonumber\\
& = &
R(\flat(V) \hookrightarrow V)(\breve{\alpha}'_{V}(q^{\flat})) \nonumber\\
& = &
\breve{\alpha}'_{\flat(V)}(q^{\flat}) \nonumber\\
& = &
\alpha_{\flat(V)}(q^{\flat}) \nonumber\\
& = &
(\flat^{*}\alpha)_{\flat(V)}(q^{\flat}) \nonumber\\
& = &
(\flat^{*}\alpha)_{V}(q^{\flat}) ,
\end{eqnarray}
which implies the commutativity of diagram (\ref{eq:sheafification:uniqueness}).
\hspace{\fill}$\square$

\subsection{Topos conditions}
\label{subsec:Topos conditions}

In this subsection, we check that the category $\Sh_{j} \Vboldhat$ is a topos.
Recall that a topos is a category which has 
(T1) all finite limits, (T2) power objects, and (T3) a subobject classifier.
What we do in this section is, therefore, to show that $\Sh_{j}\Vbold$ has all of (T1)-(T3).
(That the collection of all $j$-sheaves is a topos is a well-known fact~\cite{MM92}.
Any proposition in this subsection except for Proposition \ref{prp:downarrowdense} 
is true in the generalized setting of topos theory.)

\begin{prp}
 $\Sh_{j}\Vboldhat$ is finitely complete;
that is, it has all finite limits.  
\end{prp}
{\it Proof.}
This follows immediately from the fact 
the left exactness of the associated sheaf functor $\flat^{*}$.
That is, we can always calculate a finite limit in $\Sh_{j}\Vboldhat$
as that in $\Set^{\Vboldop}$;
the latter equals the former because $\flat^{*}$ preserves
finite limits.
\hspace{\fill}$\square$

Also, a power object of sheaves can be calculated 
as that of presheaves.
More precisely, the following proposition is well-known:
\begin{prp}
\label{prp:power}
In general, if $Q$ is a presheaf and $R$ is a sheaf,
their power object $R^{Q}$ in $\Set^{\Vboldop}$ is a sheaf.
If $Q$, also, is a sheaf, then $R^{Q}$ is a power object in $\Sh_{j}\Vboldhat$;
that is, for every sheaf $S$,
\begin{equation}
\Hom_{\Sh_{j}\Vboldhat}(S \times Q, R) \simeq \Hom_{\Sh_{j}\Vboldhat}(S , R^{Q}).
\label{eq:powerhom}
\end{equation}
\end{prp}

In order to verify this proposition,
recall that the power object $R^{Q}$ in $\Set^{\Vboldop}$ is
given by
\begin{equation}
R^{Q}(V) 
=
\Hom_{\Vboldhat}(Q_{\downarrow V}, R),
\label{eq:powerobj}
\end{equation}
where, $Q_{\downarrow V}$ is a subobject of $Q$
defined by
\begin{equation}
Q_{\downarrow V}(V') :=
\begin{cases}
Q(V') & (V' \subseteq V), \\
\emptyset & (V' \not\subseteq V). \\
\end{cases}
\label{eq:Qdownarrowobj}
\end{equation}
We should note the following property:
\begin{prp}
\label{prp:downarrowdense}
For all $Q \in \Set^{\Vboldop}$,
$Q_{\downarrow \flat(V)}$ is a dense subobject of $Q_{\downarrow V}$.
\end{prp}
{\it Proof.}
Since $Q_{\downarrow \flat(V)} \subseteq Q_{\downarrow V}$,
we have
$\overline{Q_{\downarrow \flat(V)}} \subseteq Q_{\downarrow V}$,
whereas, for each $V' \subseteq V$,
we have
\begin{eqnarray}
q' \in Q_{\downarrow V}(V')
& \implies &
q'|_{\flat(V')} \in Q_{\downarrow V}(\flat(V')) \nonumber\\
& \iff &
q'|_{\flat(V')} \in Q(\flat(V')) \nonumber\\
& \iff &
q'|_{\flat(V')} \in Q_{\downarrow \flat(V)}(\flat(V')) \nonumber\\
& \iff &
q' \in \overline{Q_{\downarrow \flat(V)}}(V'),
\end{eqnarray}
namely,
$Q_{\downarrow V} \subseteq \overline{Q_{\downarrow \flat(V)}}$.
Thus, that $Q_{\downarrow V} = \overline{Q_{\downarrow \flat(V)}}$ results.
\hspace{\fill}$\square$

{\it Proof of Proposition \ref{prp:power}.}
In order to prove the first half statement,
we need to show $\xymatrix{R^{Q} \ar [r] ^-{\zeta_{R^{Q}}} & \flat^{*}(R^{Q})}$
is an isomorphism.
Note that we have
\begin{equation}
\flat^{*}(R^{Q})(V) 
=
R^{Q}(\flat(V)) 
=
\Hom_{\Vboldhat}(Q_{\downarrow  \flat(V)}, R).
\end{equation}
From the definition of $\zeta_{R^{Q}}$,
for each $\alpha \in R^{Q}(V)$,
the morphism $(\zeta_{R^{Q}})_{V}(\alpha)=\alpha|_{\flat(V)} \in \flat^{*}(R^{Q})(V)$
is given by the commutative diagram 
\begin{equation}
\xymatrix{
Q_{\downarrow V} \ar [rr] ^{\alpha} && R \\
&& \\
\;Q_{\downarrow \flat(V)} \,.\ar @{^{(}->} [uu] ^{\iota} \ar [rruu] _{\alpha|_{\flat(V)}} && \\
}
\end{equation}
As previously shown, however, $Q_{\downarrow \flat(V)}$ is
a dense subobject of $Q_{\downarrow V}$
and $R$ is a sheaf.
Therefore, it readily follows that $(\zeta_{R^{Q}})_{V}:\alpha \mapsto \alpha|_{\flat(V)}$
is a bijection.
Thus, the first half statement is verified.
The last half statement, equation (\ref{eq:powerhom}), is obvious since $\Sh_{j}\Vboldhat$ is a full
subcategory of $\Set^{\Vboldop}$. \hspace{\fill}$\square$


The rest of this subsection is devoted to
derive the subobject classifier $\Omega_{j}$ of $\Sh_{j} \Vboldhat$.
First, we note the following fact:
\begin{prp}
Let $R$ be a quantization sheaf.
Then, every subobject $S$ of $R$ is a quantization sheaf 
if and only if $S$ is closed in $R$
(i.e., $\overline{S} \simeq S$).
\end{prp}
{\it Proof.}
To show this, it is sufficient to consider the case
where $S \subseteq \flat^{*}R$.
Since, for each $V \in \Vbold$,
we have
\begin{eqnarray}
\overline{S}(V)
& = &
\{
r^{\flat} \in \flat^{*}R(V ) \,|\, r^{\flat}||_{\flat(V)} \in S(\flat(V) ) 
\} \nonumber\\
& = &
\{
r^{\flat} \in R(\flat(V ) ) \,|\, r^{\flat}|_{\flat\flat(V)} \in S(\flat(V) ) 
\} \nonumber\\
& = &
\{
r^{\flat} \in R(\flat(V ) ) \,|\, r^{\flat}\in S(\flat(V) ) 
\} \nonumber\\
& = &
S(\flat(V)),
\end{eqnarray}
namely,
\begin{equation}
\overline{S} = \flat^{*}S.
\end{equation}
The statement of the above proposition immediately follows from this equality. 
\hspace{\fill}$\square$

From diagram (\ref{eq:closure-j}),
we have
\begin{equation}
\overline{S} = S \iff j \circ \chi = \chi ,
\end{equation}
where $\chi$ is the characteristic morphism corresponding to $S \hookrightarrow R$.

From the above-mentioned properties,
we can conclude that
\begin{equation}
\Sub_{\Sh_{j}\Vboldhat}(R)
=
{\rm Cl}\Sub_{\Vboldhat} (R)
\simeq 
\{
\chi \in \Hom_{\Vboldhat}(R, \Omega) \,|\, j \circ \chi = \chi
\}
\label{eq:subsheaf-closed-char}
\end{equation}
where, $\Sub_{\Sh_{j}\Vboldhat}(R)$ is a collection of subsheaves of $R$
(that is, subobjects of $R$ in $\Sh_{j}\Vboldhat$)
and ${\rm Cl}\Sub_{\Vboldhat} (R)$ is that of closed subobject
of $R$ in $\Set^{\Vboldop}$.
At the same time, we have
the following relationship:
\begin{equation}
\{
\chi \in \Hom_{\Vboldhat}(R, \Omega) \,|\, j \circ \chi = \chi
\}
\simeq
\Hom_{\Vboldhat}(R, \Omega_{j}).
\end{equation}
Here, the object $\Omega_{j}$ is an equalizer of $\Omega \xrightarrow{j} \Omega$
and $\Omega \xrightarrow{1_{\Omega}} \Omega$;
that is, 
$\Omega_{j}$ is a subobject of $\Omega$ such that 
the monomorphism $\xymatrix{\Omega_{j} \ar @{>->} [r] ^{m}& \Omega}$ 
satisfies $j \circ m = m$ and, for every morphism $\xymatrix{Q \ar [r] ^{\alpha} & \Omega}$
satisfying $j \circ \alpha = \alpha$,
there exists a unique morphism which makes 
the triangle on the diagram
\begin{equation}
\xymatrix{
\Omega_{j}	\ar@{>->}[rr]^{m} && \Omega \ar@<1mm>[rr]^{j} \ar@<-1mm>[rr]_{1_{\Omega}} && \Omega  \\
&&& \\
Q \ar[uu]^{\alphabreve} \ar[rruu]_{\alpha} && &            
}
\end{equation}
commute.
Thus, it turns out that subsheaves of any sheaf is
classified by $\Omega_{j}$ in the presheaf topos $\Set^{\Vboldop}$.
As will be shown below, however, 
$\Omega_{j}$ is in fact a sheaf.
Therefore, we have the following result:
\begin{prp}
\label{prp:Omegaj}
The subobject classifier $\Omega_{j}$ of $\Sh_{j}\Vboldhat$
is given by the equalizer of  $\xymatrix{\Omega \ar [r] ^{j} & \Omega}$
and $\xymatrix{\Omega \ar [r] ^{j} & \Omega }$ in $\Set^{\Vboldop}$
\end{prp}

In order to prove that,
we need the explicit expression of $\Omega_{j}$ as the equalizer
in  $\Set^{\Vboldop}$:
\begin{eqnarray}
\Omega_{j}(V)
& = &
\{ \omega \in \Omega(V) \,|\,
j_{V}(\omega)=\omega  \} \nonumber\\
& = & 
\{ \omega \in \Omega(V) \,|\,  \forall V' \in \Sub(V) \,(\flat (V') \in \omega \Rightarrow V' \in \omega) \}
\end{eqnarray}
and 
\begin{equation}
\Omega_{j}(V' \hookrightarrow V) := \Omega(V' \hookrightarrow V)|_{\Omega_{j}(V)}.
\end{equation}
It is routine to show that the presheaf $\Omega_{j}$ defined above
is indeed an equalizer of $j$ and $1_{\Omega}$.

{\it Proof of Proposition \ref{prp:Omegaj}.}
Note that
$\flat^{*} \Omega_{j}$ is a subobject of $\flat^{*} \Omega$ and is given by
\begin{eqnarray}
\flat^{*} \Omega_{j} (V) 
& = &\Omega_{j}(\flat(V)) \nonumber\\
& = &
\{ \omega^{\flat} \in \Omega(\flat(V)) \,|\, 
\forall V' \subseteq \flat(V)             \nonumber\\
& & \qquad
 \, (\flat(V') \in 
\omega^{\flat} \Rightarrow V' \in \omega^{\flat}) \} .
\end{eqnarray}
In order to see that
the morphism $\xymatrix{\Omega_{j} \ar [r]^{\zeta_{\Omega_{j}}} & \flat^{*} \Omega_{j}}$
is an isomorphism,
let us define a map
$\varrho_{V}:\flat^{*} \Omega_{j}(V) \to \Omega_{j}(V)$ by
\begin{equation}
\varrho_{V}(\omega^{\flat}) 
:= 
\{ V' \in \Sub(V) \,|\,  \flat(V') \in \omega^{\flat}\}.
\end{equation}
In fact, it is well-defined because
for all $V' \in \Sub(V)$, 
if $\flat(V') \in \varrho_{V}(\omega^{\flat}) $,
then  $\flat(\flat(V')) = \flat(V') \in \omega^{\flat}$,
hence, $V' \in \varrho_{V}(\omega^{\flat})$.
Therefore, $\varrho_{V}(\omega^{\flat}) \in \Omega_{j}(V)$.

The maps $(\zeta_{\Omega_{j}})_{V}$ and $\varrho_{V}$ are mutually inverse.
For each $\omega \in \Omega_{j}(V)$, we have
\begin{eqnarray}
\varrho_{V}((\zeta_{\Omega_{j}})_{V}(\omega))
& = &
\varrho(\omega|_{\flat(V)}) \nonumber\\
& = &
\{
V' \in \Sub(V) \,|\,
\flat(V') \in \omega|_{\flat(V)} 
\}  \nonumber\\
& = &
\{
V' \in \Sub(V) \,|\,
\flat(V') \in \omega
\} \nonumber\\
& = &
\{
V' \in \Sub(V) \,|\,
V' \in \omega
\} \nonumber\\
& = &
\omega.
\end{eqnarray}
Conversely, for $\omega^{\flat} \in \flat^{*}\Omega_{j}(V)$,
we have
\begin{eqnarray}
(\zeta_{\Omega_{j}})_{V}(\varrho_{V}(\omega^{\flat}))
& = &
\{ V'  \in \Sub(V) \,|\, \flat(V') \in \omega^{\flat}\}|_{\flat(V)} \nonumber\\
& = &
\{ V'  \in \Sub(\flat(V)) \,|\, \flat(V') \in \omega^{\flat}\} \nonumber\\
& = &
\{ V'  \in \Sub(\flat(V)) \,|\, V'\in \omega^{\flat}\} \nonumber\\
& = &
\omega^{\flat} .
\end{eqnarray}
As the inverse of $(\zeta_{\Omega_{j}})_{V}$,
$\varrho_{V}$ can be lifted up to a natural transformations $\varrho$.
We, thus, verified that $\zeta_{\Omega_{j}}$ is an isomorphism.
\hspace{\fill}$\square$

\begin{dfn}
We call  the topos $\Sh_{j}\Vboldhat$ the quantization sheaf topos
induced by the quantization topology $j$.
\end{dfn}

\section{Prequantization Topoi and Quantum Sheaf Topos}
\label{sec:Prequantization Topoi and}

We return to the setting in 
\S\ref{sec:Prequantization Categories}-\S\ref{sec:Topologies on Quantum} again.
That is, throughout this section,
we suppose that we are given fixed prequantization categories $\Cbold$
with a fixed proper prequantization category $\Abold$ which
induces a quantization topology $j$ via a quantization functor $\phi$
and a classicization functor $\psi$.
Our purpose in this section is to overview 
the interrelationships among the prequantization topoi 
$\Set^{\Cboldop}$'s and $\Set^{\Aboldop}$
and the quantization sheaf topos $\Sh_{j}\Vboldhat$.

We start with several preparatory propositions.
\begin{prp}
\label{prp:etaphiPiso}
Let $\Cbold \in \Cmath$.
Then, for all $P \in \Set^{\Cboldop},$
$\xymatrix{\phi_{*}P \ar [r] ^-{\eta_{\phi_{*}P}} & \phi_{*}\phi^{*}\phi_{*}P}$
is a natural isomorphism.
\end{prp}
{\it Proof.}
The morphism $\eta_{\phi_{*}P}$ can be calculated from definition (\ref{eq:eta}).
The inverse $\eta^{-1}_{\phi_{*}P}$ is given by the following definition:
For each $V \in \Vbold$,
the $V$-component map $(\eta^{-1}_{\phi_{*}P})_{V}:\phi_{*}\phi^{*}\phi_{*}P(V) \to \phi_{*}P(V)$
assigns, to each natural transformation 
$\xymatrix{(1_{\downarrow V})_{\phi} \ar [r] ^{\alpha} & (\phi_{*}P)_{\phi}} \in \phi_{*}\phi^{*}\phi_{*}P(V)$,
a natural transformation $\xymatrix{(1_{\downarrow V})_{\phi} \ar [rr] ^-{(\eta^{-1}_{\phi_{*}P})_{V}(\alpha)} && P}$
whose $C \in \Cbold$ component, the map $(\eta^{-1}_{\phi_{*}P})_{V}(\alpha)_{C}:(1_{\downarrow V})(\phi(C)) \to P(C)$ 
is given by
\begin{equation}
(\eta^{-1}_{\phi_{*}P})_{V}(\alpha)_{C}(\pt) = (\alpha_{C}(\pt))_{C}(\pt).
\end{equation}
Here, note that, since
$\alpha_{C}$ is a map from $(1_{\downarrow V})(\phi(C)) $ to $(\phi_{*}P)(\phi(C))$,
hence $\alpha_{C}(\pt) \in \Hom_{\Cboldhat}((1_{\downarrow V})_{\phi} ,P)$,
$(\alpha_{C}(\pt))_{C}$ is a map from $(1_{\downarrow V})(\phi(C)) $ to $P(C)$.
Therefore, $(\alpha_{C}(\pt))_{C}(\pt) \in P(C)$.
\hspace{\fill}$\square$

As was noted in \S\ref{subsec:Proper prequantization category},
if $\Cbold \in \Cmath$ includes the proper quantization category $\Abold$,
we can define the classicization functor $\psi:\Vbold \to \Cbold$ 
from the inverse of the quantization map $\uptilde$.
This induces the geometric functor 
$\psi= (\psi^{*}, \psi_{*}):\Set^{\Vboldop} \to \Set^{\Cboldop}$.
The following relation, which relates $\phi_{*}$
and $\psi^{*}$ is very useful for us:
\begin{prp}
\label{prp:phivarpipsi}
Suppose that $\Cbold \in \Cmath$ includes 
the proper prequantization category $\Abold$
as a subcategory.
Then, there exists a natural isomorphism
$\xymatrix{\phi_{*} \ar [r] ^{\varpi} _{\sim} & \psi^{*}}$.
Here, $\phi_{*}$ and $\psi^{*}$
are the direct image part of the geometric morphism 
$\phi:\Set^{\Cboldop} \to \Set^{\Vboldop} $
and the inverse image part of  
$\psi:\Set^{\Vboldop} \to \Set^{\Cboldop}$,
respectively.
\end{prp}
{\it Proof.}
To prove this,
let us recall that $\phi_{*} P (V)$
is given by (\ref{eq:phi*P(V)}) for every $V \in \Vbold$.
It follows that
\begin{eqnarray}
(1_{\downarrow V})_{\phi}(C) \not = \emptyset
& \iff &
\phi(C) \subseteq V \nonumber\\
& \iff &
C \subseteq \psi(V) .
\end{eqnarray}
Also, note that $\psi(V) \in \Cbold$ since $\Abold \subseteq \Cbold$.
We therefore have a bijection $(\varpi_{P})_{V}:\phi_{*}P(V) \to \psi^{*}P(V)$
defined by
\begin{equation}
(\varpi_{P})_{V}(\alpha):=\alpha_{\psi (V)}(\pt) \in P(\psi(V))=\psi^{*}P(V),
\end{equation}
for each natural transformation
$\xymatrix{(1_{\downarrow V})_{\phi} \ar [r] ^{\alpha} & P} \in \phi_{*}P(V)$.
Here, the inverse map $(\varpi^{-1}_{P})_{V}:{\psi}^{*}P(V) \to \phi_{*}P(V)$ of $(\varpi_{P})_{V}$ 
is defined as follows;
for each $p^{\psi} \in P_{\psi}(V)$, $(\varpi^{-1}_{P})_{V}$ assigns a natural transformation 
$(1_{\downarrow V})_{\phi} \xrightarrow{(\varpi^{-1}_{P})_{V}(p^{\psi})} P$ given by

\begin{equation}
(\varpi^{-1}_{P})_{V}(p)_{C}(\pt) := P(C \hookrightarrow \psi (V))(p^{\psi}) \equiv p^{\psi}|_{C}
\end{equation}
for each $C \subseteq \psi(V)$.
Furthermore, for each $V' \hookrightarrow V$,
it follows that
\begin{eqnarray}
P_{\psi}(V' \hookrightarrow V)((\varpi_{P})_{V}(\alpha))
& = &
P_{\psi}(V' \hookrightarrow V)(\alpha_{\psi (V)}(\pt)) \nonumber\\
& = &
(\alpha_{\psi (V)}(\pt) )||_{V'}\nonumber\\
& = &
(\alpha_{\psi (V)}(\pt) )|_{\psi(V')}\nonumber\\
& = &
\alpha_{\psi (V')}(\pt) \nonumber\\
& = &
(\alpha|_{V'})_{\psi (V')}(\pt) \nonumber\\
& = &
(\varpi_{P})_{V'} (\alpha|_{V'})  \nonumber\\
& = &
(\varpi_{P})_{V'} (\phi_{*}P (V' \hookrightarrow V) (\alpha)),
\end{eqnarray} 
hence, the naturality diagram
\begin{equation}
\xymatrix{
\phi_{*}P(V) \ar [rr] ^{(\varpi_{P})_{V}} _{\sim} \ar [dd] _{\phi_{*}P(V' \hookrightarrow V)} 
            & & {\psi}^{*}P(V) \ar [dd] ^{\psi^{*}P(V' \hookrightarrow V)} \\
            && \\
\phi_{*}P(V') \ar [rr] _{(\varpi_{P})_{V'}} ^{\sim} & & {\psi}^{*}P(V') \\
}
\end{equation}
of $(\varpi_{P})_{V}$ with respect to $V$ commutes.
Thus, we obtain the natural isomorphism 
$\xymatrix{\phi_{*} P \ar [r] ^{\varpi_{P}}_{\sim} & \psi^{*}P}$. 

Next, we need to show that $\varpi_{P}$ is natural for $P$,
namely, that the diagram
\begin{equation}
\xymatrix{
\phi_{*}P(V) \ar [rr] ^{(\varpi_{P})_{V}} _{\sim} \ar [dd] _{\phi_{*}(P \xrightarrow{\theta} P')_{V}}
            & & {\psi}^{*}P(V) \ar [dd] ^{\psi^{*}(P \xrightarrow{\theta} P')_{V}} \\
            && \\
\phi_{*}P'(V) \ar [rr] _{(\varpi_{P'})_{V}} ^{\sim} & & {\psi}^{*}P'(V) \\
}
\end{equation}
commutes for every morphism $\xymatrix{P \ar [r] ^{\theta} & P'} \in \Mor(\Set^{\Cboldop})$.
This is shown as follows:
\begin{eqnarray}
\psi^{*}(P \xrightarrow{\theta} P') \circ (\varpi_{P})_{V}(\alpha)
& = & 
\psi^{*}(P \xrightarrow{\theta} P') (\alpha_{\psi(V)}(\pt)) \nonumber\\
& = &
\theta_{\psi(P)}(\alpha_{\psi(V)}(\pt)) \nonumber\\
& = &
(\theta \circ \alpha)_{\psi(V)}(\pt) \nonumber\\
& = &
(\varpi_{P'})_{V}(\theta \circ \alpha) \nonumber\\
& = &
(\varpi_{P'})_{V} \circ \phi_{*}(P \xrightarrow{\theta} P')(\alpha).
\end{eqnarray}
\hspace{\fill}$\square$

From Proposition \ref{prp:phivarpipsi}, we can readily verify the following proposition:
\begin{prp}
Suppose the same condition as Proposition \ref{prp:phivarpipsi}.
Then, we have a natural isomorphism
\begin{equation}
\phi_{*}\phi^{*} \xrightarrow[\sim]{\varpi_{\phi^{*}}} \flat^{*}.
\label{eq:phiphi->flat}
\end{equation}
\end{prp}
{\it Proof.}
Note that, for every $Q \in \Set^{\Vboldop}$,
it follows that
\begin{equation}
\flat^{*} Q
=
Q_{\flat} \nonumber\\
=
Q_{\phi \circ \psi} \nonumber\\
=
(\phi^{*}Q)_{\psi} \nonumber\\
=
\psi^{*}\phi^{*} Q .
\end{equation}
Therefore, from Proposition \ref{prp:phivarpipsi},
we have a natural morphism $\phi_{*}\phi^{*}Q \xrightarrow[\sim]{\varpi_{\phi^{*}Q}} \flat^{*}Q$.
\hspace{\fill}$\square$

\begin{prp}
\label{prp:zetavarpieta}
Suppose the same condition as Proposition \ref{prp:phivarpipsi}.
Then, the diagram 
\begin{equation}
\xymatrix{
&&  \flat^{*} \\
I \ar [rru] ^{\zeta} \ar [rrd] _{\eta} && \\
&&  \phi_{*}\phi^{*} \ar [uu] _{\varpi_{\phi^{*}}}^{\wr}\\
}
\end{equation}
commutes.
\end{prp}
{\it Proof.}
Note that for each $Q \in \Set^{\Vboldop}$,
$V \in \Vbold$, and $q \in Q(V)$,
\begin{eqnarray}
(\varpi_{\phi^{*}Q})_{V} \circ \eta_{Q(V)}(q) 
& = & (\eta_{Q(V)}(q))_{\psi(V)}(\pt) \nonumber\\
& = & q|_{\phi(\psi(V))} \nonumber\\
& = & q|_{\flat(V)}\nonumber\\
& = & \zeta_{V}(q).
\end{eqnarray}
\hspace{\fill}$\square$

\begin{prp}
\label{prp:quantumsheafexternalnecessary}
Suppose the same condition as Proposition \ref{prp:phivarpipsi}.
Then, for every quantization sheaf $R$,
there exists a presheaf $P \in \Cbold$ such that
\begin{equation}
R \simeq \phi_{*} P.
\end{equation}
\end{prp}
{\it Proof.}
Since $R$ is a quantization sheaf,
$\xymatrix{R \ar [r] ^{\zeta_{R}} & \flat^{*}R}$
is an isomorphism.
Thus, from Proposition \ref{prp:zetavarpieta},
$\xymatrix{R \ar [r] ^-{\eta_{R}} & \phi_{*}\phi^{*} R}$ is isomorphism.
Therefore, we can choose $P = \phi^{*}R$.
\hspace{\fill}$\square$

\begin{prp}
\label{prp:quantumsheafexternalsufficient}
Suppose the same condition as Proposition \ref{prp:phivarpipsi}.
Then, for all $P \in \Set^{\Cboldop}$, we have
\begin{equation}
\phi_{*} P \in \Sh_{j} \Vboldhat.
\end{equation}
\end{prp}
{\it Proof.}
According to Proposition \ref{prp:etaphiPiso}
and Proposition \ref{prp:zetavarpieta}, 
$\xymatrix{\phi_{*} P \ar [r] ^{\zeta_{\phi_{*} P}} & \flat^{*} \phi_{*}P }$
is a natural isomorphism for all $P \in \Set^{\Cboldop}$.
\hspace{\fill}$\square$

From propositions \ref{prp:quantumsheafexternalnecessary}
and \ref{prp:quantumsheafexternalsufficient},
we have another form of the sheaf condition:
\begin{thm}
\label{thm:quantumsheafexternalcondition}
Let $\Cbold \in \Cmath$ and $\Abold \subseteq \Cbold$.
Then, it follows that, for every $Q \in \Set^{\Vboldop}$,
\begin{equation}
Q \in \Sh_{j}\Vboldhat
\iff
\exists P \in \Set^{\Cboldop} \mbox{ s.t. } Q \simeq \phi_{*} P.
\end{equation}
\end{thm}

In order to see Theorem \ref{thm:quantumsheafexternalcondition}
from a different viewpoint,
we return to commutative diagram (\ref{eq:CAV}).
Let $\Cbold$ be an arbitrary prequantization category.
(It does not need to include $\Abold$.)
Then,
diagram (\ref{eq:CAV}) gives the commutative diagram of geometric morphisms,
\begin{equation}
\xymatrix{
\Set^{\Cboldop} \ar [rr] ^{\sharp} \ar [rdd] _{\phi} &  & \Set^{\Aboldop} \ar [ldd] ^{\phi} \\
&& \\
    & \;\; \Set^{\Vboldop}\,. & \\
},
\label{eq:ChatAhatVhat}
\end{equation}
Here, this diagram means that, for the inverse image parts,
\begin{equation}
\sharp^{*} \phi^{*}  Q = Q_{\phi \circ \sharp } = Q_{\phi} = \phi^{*} Q
\end{equation}
for all $Q\in \Set^{\Vboldop}$, and for the direct image parts,
\begin{equation}
\phi_{*} \sharp_{*} P \simeq \phi_{*} P
\label{eq:phi*sharp*Psimeqphi*P}
\end{equation}
for all $P \in \Set^{\Pboldhat}$.
Relation (\ref{eq:phi*sharp*Psimeqphi*P}) comes from the fact that,
for each $V \in \Vbold$,
\begin{eqnarray}
\phi_{*} \sharp_{*} P (V)
& = &
\Hom_{\Aboldhat} ( (1_{\downarrow V})_{\phi}, \sharp_{*} P ) \nonumber\\
& \simeq &
\Hom_{\Cboldhat} ( \sharp^{*} (1_{\downarrow V})_{\phi}, P ) \nonumber\\
& = &
\Hom_{\Cboldhat} (  (1_{\downarrow V})_{\phi \sharp}, P ) \nonumber\\
& = &
\Hom_{\Cboldhat} (  (1_{\downarrow V})_{\phi}, P ) \nonumber\\
& = &
\phi_{*}  P (V).
\end{eqnarray}

According to Theorem \ref{thm:quantumsheafexternalcondition},
the collection $\phi_{*}(\Set^{\Aboldop})$ of all of direct images $\phi_{*}(P')$ 
($P' \in \Set^{\Aboldop}$)
and all of their isomorphic objects in $\Set^{\Vboldop}$
equals the quantization sheaf topos $\Sh_{j}\Vboldhat$.
On the other hand, from equality (\ref{eq:phi*sharp*Psimeqphi*P}),
we have $\phi_{*}P \simeq \phi_{*}(\sharp_{*} P)$  for all $P \in \Set^{\Cboldop}$.
Since $\sharp_{*} P \in \Set^{\Aboldop}$,
every direct image of $P \in \Set^{\Cboldop}$ by $\phi_{*}$ is
a quantization sheaf.
Thus, we can say that the collection $\phi_{*}(\Set^{\Cboldop})$
is a part of the quantization sheaf topos $\Sh_{j} \Vboldhat$.

If, furthermore, $\Cbold$ includes $\Abold$ as a subcategory,
we have the commutative diagram,
\begin{equation}
\xymatrix{
\Set^{\Cboldop}  \ar [rdd] _{\phi} &  & \Set^{\Aboldop} \ar [ll] _{\iota} \ar [ldd] ^{\phi} \\
&& \\
    & \;\; \Set^{\Vboldop} \,, & \\
}
\label{eq:AhatChatVhat}
\end{equation}
from diagram (\ref{eq:C'CV}).
This diagram implies that, 
for all $P' \in \Set^{\Aboldop}$,
there exists $P \in \Set^{\Cboldop}$ such that
$\phi_{*}P' \simeq \phi_{*}P.$.
Therefore,
if $\Cbold$ includes $\Abold$,
$\phi_{*}(\Set^{\Cboldop})$ covers all of the quantization sheaf topos $\Sh_{j} \Vboldhat$;
that is, $\phi_{*}(\Set^{\Cboldop}) = \Sh_{j} \Vboldhat$.
This is what Theorem \ref{thm:quantumsheafexternalcondition} states.

Finally, we should note that the results in this section
are a special case of the well-known factorization theorems 
for geometric morphisms~\cite{MM92}.

\section{Concluding Remarks}
\label{sec:Concluding Remarks}

In the present paper, 
we clarified the structures which quantization induces in the quantum topos;
they are the quantization topology and the quantization sheaf topos.
We showed that
any quantization map $\upsilon$ induces a quantization topology $j$ 
(or its equivalent, a quantization Grothendieck topology $J$) 
on the quantum topos $\Set^{\Vboldop}$.
It is given by means of the endofunctor $\flat:= \phi\psi  :\Vbold \to \Vbold$,
where $\phi$ and $\psi$ are the quantization functor and the classicization functor,
respectively,  
between $\Vbold(\Hcal)$ and an arbitrary prequantization
category $\Cbold$ including the proper prequantization category $\Abold$.
 We, further, clearly specified the condition
for a presheaf in $\Set^{\Vboldop}$ to be a quantization sheaf, in different two ways.
A presheaf $Q$ is a quantization sheaf if and only if 
the morphism $\xymatrix{Q \ar [r] ^{\zeta_{Q}} & \flat^{*}Q}$ is an isomorphism.
Indeed, $\flat^{*}$ is the associated sheaf functor to the quantization sheaf topos.
Also, a presheaf is a quantization sheaf if and only if it is isomorphic to
a direct image of an object of $\Set^{\Cboldop}$
where $\Cbold$ is a prequantization category including $\Abold$.
Since any prequantization category including $\Abold$
induces the quantization topology, it induces the quantization sheaf topos.

However, there still remain issues we should address at the present elementary level. 
In this paper,
we assumed that the quantization map $\upsilon$ is fixed.
We can, however, adopt unitarily equivalent quantization map.
Since unitarily equivalent quantization 
gives physically equivalent theory,
the quantization topology and the sheaf topos
have to be transformed in a `covariant' way
in order for them to be meaningful.
It is easy to see that the proper quantization
category $\Abold$ is invariant under unitary equivalence
of quantization.
We, however, 
have to establish the notion of covariance appropriately 
for topologies and sheaves.
Also, it would be important to clarify
how the irreducibility of quantization map works.
In fact, in the present paper, the assumption of irreducibility for quantization
does not work;
our theory is applicable to both of reducible and irreducible 
representations of classical observables.
Irreducibility might put some restrictions on possible forms of quantum topologies.

We also neglected applications to theories for physics.
We expect that we can develop quantum theory
on the quantization topos $\Sh_{j}\Vboldhat$
as D\"{o}ring and Isham did~\cite{DI08a,DI08b,DI08c}.
In fact,
we can easily derive a lattice homomorphism
from $\Omega$ to $\Omega_{j}$ from 
the quantization topology $j$.
Though it is not a Heyting algebra homomorphism because
it does not preserve the pseudo-compliment,
we expect that physically meaningful propositions 
independent of the pseudo-complimen can be 
coded on the quantization sheaf topos 
via the associated sheaf functor $\flat^{*}$.
Also, we expect that 
sophisticated ingredients in topos quantum theory,
such as the `daseinization' operator and the quantity value object~\cite{DI08b, DI08c},
can be defined on the quantization sheaf topos.
Concerning classical physics,
we have prequantization topoi where we can code classical physics,
though the best topos for classical physics would be $\Set$~\cite{DI08a}. 
We expect this can be translated to $\Sh_{j}\Vboldhat$ by $\phi_{*}$.
If that is the case,
classical physics and quantum physics might be 
coded in the same logical space $\Omega_{j}$.

We hope to return to some of the above-mentioned issues in subsequent papers.

\section*{Acknowledgement}
This work was done while I was visiting Oxford University Computing Laboratory 
(Department of Computer Science) from September 2010 to July 2011
as an overseas research personnel of Ryukoku University.
I am grateful to Dr. A. D\"{o}ring for his kind hospitality at Oxford.

\appendix

\section{Adjunction $\phi^{*} \dashv \phi_{*}$}
\label{subsec:Adjunction}

In this appendix,
we prove that the pair $(\phi^{*},\phi_{*})$ defined in \S\ref{sec:Geometric Morphisms}
is an adjunction, namely, $\phi^{*} \dashv \phi_{*}$.
In order to do so,
it is sufficient to show that
there exists a natural transformation
$\xymatrix{1 \ar [r] ^-{\eta} & \phi_{*} \phi^{*} }$, the unit of adjunction,
such that,
for any morphism $\xymatrix{Q \ar [r] ^-{\tau} & \phi_{*} P}$ in $\Set^{\Vboldop}$,
there exists a unique morphism 
$\xymatrix{\phi^{\ast}Q \ar [r] ^-{\sigma} & P}$ in $\Set^{\Cboldop}$
which makes the diagram 
%
%

\begin{equation}
	\xymatrix{
Q	\ar[rr]^{\tau} \ar[rrdd]_{\eta_{Q}} && \phi_{\ast}P  \\
&& \\
	&& \phi_{\ast}\phi^{\ast} Q  \ar [uu]_{\phi_{\ast} \sigma}\\
	}
	\label{eq:tausigmaeta}
\end{equation}
commute.

Before giving the definition of $\eta$ 
and constructing $\sigma$,
we need to mention a few points.
First, note that the naturality condition for $\tau$,
the commutatibity of the diagram
\begin{equation}
\xymatrix{
Q(V) \ar [dd] _{Q(V' \hookrightarrow V)} \ar [rr] ^{\tau_{V}} && \phi_{*}P(V) \ar [dd] ^{\phi_{*}P(V' \hookrightarrow V)} \\
&& \\
Q(V') \ar [rr] _{\tau_{V'}} && \phi_{*}P(V') \\
}
\end{equation}
is equivalent to that of the diagram
\begin{equation}
\xymatrix{
\phi^{*}(1_{\downarrow V}) \ar [rr] ^{\tau_{V}(q)} && P \\
&& \\
\phi^{*}(1_{\downarrow V'}) \ar [uu] ^{\phi^{*}(1_{\downarrow (V' \hookrightarrow V)})} 
\ar [rruu] _{\tau_{V'}(q|_{V'})} &&  \\
}
\end{equation}
for all $q \in Q(V)$.
This is, furthermore, equivalent to 
\begin{equation}
\tau_{V}(q)_{C}(\pt) = \tau_{V'}(q|_{V'})_{C}(\pt) 
\label{eq:taucondition1}
\end{equation}
for all $C \in \Cbold$ such that $\phi(C) \subseteq V$.

Also, since each $\tau_{V}(q)$ ($q \in Q(V)$) is a morphism of $\Set^{\Cboldop}$,
it has to make the diagram
\begin{equation}
\xymatrix{
\phi^{*}(1_{\downarrow V})(C) \ar [rr] ^-{\tau_{V}(q)_{C}} 
\ar [dd] _{\phi^{*}(1_{\downarrow V})(C' \hookrightarrow C)}
 && P(C) \ar [dd] ^{P(C' \hookrightarrow C)}\\
&& \\
\phi^{*}(1_{\downarrow V})(C') \ar [rr] _-{\tau_{V}(q)_{C'}} && P(C') \\
}
\end{equation}
commute as the naturality condition.
This is, further, equivalent to
\begin{equation}
\tau_{V}(q)_{C}(\pt )|_{C'} = \tau_{V}(q)_{C'}(\pt)
\label{eq:taucondition2}
\end{equation}
for all $C' \hookrightarrow C \in \Mor(\Set^{\Cboldop})$
such that $\phi(C) \subseteq V$.
Thus, it turns out that every $\tau \in \Hom_{\Vboldhat}(Q, \phi_{*}P)$
is determined by the data $\tau_{V} (q) _{C}(\pt)$ with $q \in Q(V)$ and $\phi(C) \subseteq V$
satisfying (\ref{eq:taucondition1}) and (\ref{eq:taucondition2}).

Since $\eta_{Q} \in \Hom_{\Vboldhat}(Q, \phi_{*}(\phi^{*}Q))$,
$\eta_{Q}$ is determined by data $(\eta_{Q})_{V}(q)_{C}(\pt)$ with $q \in Q(V)$ and $\phi(C) \subseteq V$
satisfying conditions corresponding to  (\ref{eq:taucondition1}) and (\ref{eq:taucondition2}),
namely, the naturality of $\eta_{Q}$,
\begin{equation}
(\eta_{Q})_{V}(q)_{C}(\pt) = (\eta_{Q})_{V'}(q|_{V'})_{C}(\pt) ,
\label{eq:etacondition1}
\end{equation}
and the naturality of $(\eta_{Q})_{V}(q)$,
\begin{equation}
(\eta_{Q})_{V}(q)_{C}(\pt)||_{C'} = (\eta_{Q})_{V}(q)_{C'}(\pt).
\label{eq:etacondition2}
\end{equation}
Now, we define $\eta_{Q}$ by
\begin{equation}
(\eta_{Q})_{V}(q)_{C}(\pt) :=  q|_{\phi(C)} \in Q_{\phi}(V).
\label{eq:eta}
\end{equation}
It is easy to check that definition \ref{eq:eta} satisfies 
(\ref{eq:etacondition1}) and (\ref{eq:etacondition2}).
Also, the naturality condition for $\eta$,
the commutativity of the diagram
\begin{equation}
\xymatrix{
Q \ar [rr] ^{\eta_{Q}} \ar [dd] _{\theta} && \phi_{*}\phi^{*}Q \ar [dd] ^{\phi_{*}\phi^{*}\theta} \\
&& \\
Q' \ar [rr] ^{\eta_{Q'}} && \phi_{*}\phi^{*}Q'  \\
}
\end{equation}
is easily proved with aid of the naturality of $\theta$.

For an arbitrary $\tau$,
we define a natural transformation $\xymatrix{Q_{\phi} \ar [r] ^-{\sigma} & P} $ 
by giving each $\sigma_{C}:Q_{\phi}(C) \to  P(C)$ by
\begin{equation}
\sigma_{C} (q) := \tau_{\phi(C)}(q)_{C}(\pt).
\end{equation}
This definition satisfies the naturality condition
\begin{equation}
\xymatrix{
Q_{\phi}(C) \ar [rr]  ^{\sigma_{C}} \ar [dd] _{Q_{\phi}(C' \hookrightarrow C)} && P(C) \ar [dd] ^{P(C' \hookrightarrow C) \;\;\;} \\
&& \\
Q_{\phi}(C') \ar [rr]  _{\sigma_{C'}} && P(C') \;,\\
}
\end{equation}
because, for any $q^{\phi} \in Q_{\phi}(C)$, it follows that
\begin{eqnarray} 
\sigma_{C}(q^{\phi})|_{C'}
& = & 
\tau_{\phi(C)}(q^{\phi})_{C}(\pt) |_{C'}  \nonumber\\
& = &
\tau_{\phi(C)}(q^{\phi})_{C'}(\pt) \nonumber\\
& = &
\tau_{\phi(C')}(q^{\phi}|_{\phi(C')})_{C'} (\pt)  \nonumber\\
& = &
\tau_{\phi(C')}(q^{\phi}||_{C'})_{C'} (\pt)  \nonumber\\
& = &
\sigma_{C'}(q^{\phi}||_{C'}) .
\end{eqnarray}

The commutativity of diagram (\ref{eq:tausigmaeta}) is equivalent to
that of diagram
\begin{equation}
\xymatrix{
(1_{\downarrow V})_{\phi}(C)  \ar [rr] ^-{(\eta_{Q})_{V}(q)_{C}} \ar [rrdd] _{\tau_{V}(q)_{C}} &  & Q_{\phi}(C) \ar [dd] ^{\sigma_{C}}\\
&& \\
                                        &  & P(C) \\
}
\label{eq:tausigmaeta2}
\end{equation}
for each $V \in \Vbold$, $q \in Q(V)$, 
and $C \in \Cbold$ such that $\phi(C) \subseteq V$.
This is proved as
\begin{eqnarray}
\sigma_{C}\circ \eta_{Q(V)}(q)_{C} (\pt)
& = &
\tau_{\phi(C)}(q|_{\phi(C)})_{C}(\pt) \nonumber\\
& = &
\tau_{V}(q)_{C}(\pt).
\end{eqnarray}

Finally, to see uniqueness of $\sigma$,
let us assume that also $\sigma'$ makes the diagram (\ref{eq:tausigmaeta})
and, hence, (\ref{eq:tausigmaeta2}) commute.
We consider the case where $V = \phi(C)$.
Then, for each $q^{\phi} \in Q_{\phi}(C)$, $q^{\phi} = q^{\phi}|_{\phi(C)}$,
and
\begin{eqnarray}
\sigma'_{C} (q^{\phi})
& = &
\sigma'_{C} (q^{\phi}|_{\phi(C)}) \nonumber\\
& = &
\sigma'_{C}\circ (\eta_{Q})_{\phi (C)}(q^{\phi})_{C} (\pt) \nonumber\\
& = &
\tau_{\phi(C)}(q^{\phi})_{C}(\pt)  \nonumber\\
& = &
\sigma_{C}(q^{\phi}).
\end{eqnarray}
Thus, $\sigma' = \sigma$ is concluded.

\section{Adjunction $\phi ! \dashv \phi^{*}$}
\label{sec:Adjunction}

In this appendix, we construct an adjunction $\phi! \dashv \phi^{*}$.

In order to do so,
first we define sets $\amalg P(V)$ for all $P \in \Set^{\Cboldop}$
and $V \in \Vbold$ by
\begin{equation}
\amalg P (V) 
:=
\coprod_{C \in \Cbold \atop (V \subseteq \phi(C))} P(C)
=
\bigcup_{C\in \Cbold} 
\{ (p,C) \,|\,V \subseteq \phi(C),\; p \in P(C) \}.
\end{equation}

We define a binary relation $\sim_{V}$ on $\amalg P(V)$;
that is, for $(p,C)$ and $(p',C') \in \amalg P(V)$,
\begin{equation}
(p,C) \sim_{V} (p',C'),
\end{equation}
if and only if
there exist a finite sequence,
$(p_{i}, C_{i})$ $(i = 1, \cdots, n)$, of elements of $\amalg P(V)$
such that 
\begin{equation}
(p,C) = (p_{1}, C_{1}), \quad (p',C') = (p_{n}, C_{n}),
\label{eq:equivalencecondition1}
\end{equation}
and for each $i = 1, \cdots, n-1$, 
there exists $\dot{C}_{i} \in \Cbold$ such that
\begin{equation}
\dot{C}_{i} \subseteq C_{i} \cap C_{i+1},
\quad
V \subseteq \phi(\dot{C}_{i}),
\quad
{\rm and}
\quad
p_{i}|_{\dot{C}_{i}} = p_{i+1}|_{\dot{C}_{i}}.
\label{eq:equivalencecondition2}
\end{equation}
It is clear that the binary relation $\sim_{V}$ is an equivalence relation.
We can therefore define the quotient space of $\amalg P(V)$ by $\sim_{V}$.
We write $\phi ! P(V)$ for it; that is,
\begin{equation}
\phi ! P(V) := \amalg P(V) / \sim_{V},
\label{eq:phi!P(V)}
\end{equation}
and $[p,C]_{V}$ for the image of $(p,C)$ by the projection $\amalg P(V) \to \phi! P(V)$.

We can define $\phi ! P(V)$ given by (\ref{eq:phi!P(V)})
as a presheaf $\phi! P \in \Set^{\Vboldop}$:
For every $V' \hookrightarrow V$,
we define a map $\phi ! P(V' \hookrightarrow V): \phi ! P (V) \to \phi ! P(V')$
by
\begin{equation}
\phi ! P(V' \hookrightarrow V) ([p,C]_{V}) :=  [p,C]_{V'} 
\label{eq:phi!ObjMor}
\end{equation} 
for all $[p,C]_{V} \in \phi ! P (V)$.
Indeed, the map given by (\ref{eq:phi!ObjMor}) is well-defined:
If $(p,C) \in \amalg P(V)$ then $(p,C) \in \amalg P(V')$ 
because $V' \subseteq V \subseteq \phi(C)$;
if $(p,C) \sim_{V} (p',C')$, we have a sequence $(p_{i}, C_{i}) \in \amalg P(V)$
accompanied with $\dot{C}_{i}$ satisfying conditions (\ref{eq:equivalencecondition1}) 
and (\ref{eq:equivalencecondition2}),
but they also induce $(p,C) \sim_{V'} (p',C')$ for $V' \subseteq V$.

Furthermore, we can regard $\phi ! P$ as a functor $\phi ! :\Set^{\Cboldop} \to \Set^{\Vboldop}$.
For each $\xymatrix{P \ar [r] ^{\theta} & P'} \in \Mor(\Set^{\Cboldop})$,
we can define a morphism $\xymatrix{\phi ! P \ar [r] ^{\phi ! \theta} & \phi ! P'} \in \Mor(\Set^{\Vboldop})$
by $V$-components $\phi ! \theta_{V} :\phi ! P(V) \to \phi ! P'(V)$
each of which is given by
\begin{equation}
\phi ! \theta_{V} ([p,C]_{V}) :=  [\theta_{C}(p), C]_{V}.
\end{equation}
Indeed, this defines a natural transformation from $\phi ! P$ to $\phi! P'$ because,
for each $V' \hookrightarrow V$, it follows that
\begin{eqnarray}
\phi ! P'(V' \hookrightarrow V)\circ \phi ! \theta_{V} ([p,C]_{V})
& = &
\phi ! P'(V' \hookrightarrow V)([\theta_{C}(p), C]_{V})  \nonumber\\
& = &
[\theta_{C}(p), C]_{V'} \nonumber\\
& = &
\phi ! \theta_{V'}([p.C]_{V'}) \nonumber \\
& = &
\phi ! \theta_{V'} \circ \phi ! P(V' \hookrightarrow V) ([p,C]_{V}) ,
\end{eqnarray}
which is the naturality condition for $\phi !$.
It is easy to see $\phi !$ preserves identity morphisms and products of morphisms.

In order to show that $\phi !$ is left adjoint to $\phi^{*}$,
we construct the unit $\xymatrix{1 \ar [r] ^-{\mu} & \phi^{*} \phi !}$.
For each $P \in \Cbold$, the $C$-component map
$(\mu_{P})_{C}: P(C) \longrightarrow \phi^{*} \phi ! P(C)$
of $ \xymatrix{P \ar [r] ^-{\mu_{P}} & \phi^{*} \phi ! P}$
is  defined by
\begin{equation}
(\mu_{P})_{C}(p) :=  [p,C]_{\phi(C)}  .
\end{equation}
In fact, this definition makes $\mu_{P}$ a natural transformation because it follows that,
for every $C' \hookrightarrow C$,
\begin{eqnarray}
\phi^{*} \phi ! P(C' \hookrightarrow C) \circ \mu_{C} (p)
& = &
\phi^{*} \phi ! P(C' \hookrightarrow C)([p,C]\phi_{C}) \nonumber \\
& = &
[p,C]||_{C'} \nonumber \\
& = &
[p,C]|_{\phi(C')} \nonumber \\
& = &
[p|_{C'},C']|_{\phi(C')} \nonumber \\
& = &
\mu_{P(C')} (p|_{C'}) \nonumber \\
& = &
\mu_{P(C')} \circ P(C' \hookrightarrow C) (p)  .
\end{eqnarray}
Also,  $\mu_{P}$ can be regarded as a morphism in $\Set^{\Cboldop}$:
For each $\xymatrix{P \ar [r] ^{\alpha} & P'}$ and $C \in \Cbold$, we have
\begin{eqnarray}
\phi^{*} \phi ! \alpha_{C} \circ \mu_{P(C)} (p)
& = &
\phi^{*} \phi ! \alpha_{C} ([p,C]_{\phi(C)}) \nonumber\\
& = &
[\alpha_{C}(p),C]_{\phi(C)} \nonumber\\
& = &
\mu_{P'(C)}(\alpha_{C}(p)) \nonumber\\
& = &
\mu_{P'(C)} \circ \alpha_{C}(p)   ,
\end{eqnarray}
for each $p \in P(C)$,
hence the naturality condition
\begin{equation}
\xymatrix{
P \ar [rr] ^{\mu_{P}} \ar [dd] _{\alpha} && \phi^{*} \phi ! P \ar [dd] ^{\phi^{*} \phi ! \alpha}\\
&& \\
P' \ar [rr] ^{\mu_{P'}} && \phi^{*} \phi ! P' \\
}      
\end{equation}
is satisfied.

For each morphism $\xymatrix{P \ar [r] ^{\nu} & \phi^{*}Q}$,
we define a morphism $\xymatrix{\phi ! P \ar [r] ^{\lambda} & Q}$ as follows:
\begin{equation}
\lambda_{V}([p,C]_{V}) := \nu_{C}(p)|_{V} .
\end{equation}
First, note that the function $\lambda_{V}$
is well defined.
In fact, if $(p, C) \sim_{V} (p', C')$, 
we have a sequence finite $(p_{i}, C_{i}) \in \phi ! P(V)$ and $\dot{C}_{i} \in \Cbold$
satisfying (\ref{eq:equivalencecondition1}) and (\ref{eq:equivalencecondition2}).
Then it follows that
\begin{eqnarray}
\nu_{C}(p)|_{V}
& = &
\nu_{C_{1}}(p_{1})|_{V} \nonumber\\
& = &
\nu_{\dot{C}_{1}}(p_{1})|_{\phi(\dot{C}_{1})}|_{V} \nonumber\\
& = &
\nu_{\dot{C}_{1}}(p_{1}||_{\dot{C}_{1}})|_{V} \nonumber\\
& = &
\nu_{\dot{C}_{1}}(p_{2}||_{\dot{C}_{1}})|_{V} \nonumber\\
& = &
\nu_{C_{2}}(p_{2})|_{\phi(\dot{C}_{1})}|_{V} \nonumber\\
& = &
\nu_{C_{2}}(p_{2})|_{V}\nonumber\\
& = &
\cdots \nonumber\\
& = &
\nu_{C_{n}}(p_{n})|_{V}  = \nu_{C'}(p')|_{V} .
\end{eqnarray}
Furthermore,
the naturality of $\lambda$ holds:
For each $V' \hookrightarrow V$,
\begin{eqnarray}
Q(V' \hookrightarrow V)\circ \lambda_{V}([p,C]_{V})
& = &
Q(V' \hookrightarrow V)(\nu_{C}(p)\_{V}) \nonumber\\
& = &
\nu_{C}(p)|_{V}|_{V'} \nonumber\\
& = &
\nu_{C}(p)|_{V'} \nonumber\\
& = &
\lambda_{V'}([p,C]_{V'}) \nonumber\\
& = &
\lambda_{V'} \circ \phi ! P(V' \hookrightarrow V)([p,C]_{V}) .
\end{eqnarray}

The commutativity of the diagram
\begin{equation}
\xymatrix{
P \ar [rr] ^{\nu} \ar [rrdd] _{\mu_{P}} && \phi^{*} Q \\
&& \\ 
&& \phi^{*} \phi ! P \ar [uu] _{\phi^{*} {\lambda}} \\
}
\label{eq:mulambdanu}
\end{equation}
follows from the fact that,
for every $C \in \Cbold$ and $p \in P(C)$,
\begin{eqnarray}
\phi^{*}\lambda_{C} \circ \mu_{P(C)} (p)
& = &
\phi^{*}\lambda_{C} ([p,C]_{\phi(C)}) \nonumber\\
& = &
\nu_{C}(P)|_{\phi(C)} \nonumber\\
& = &
\nu_{C}(P) .
\end{eqnarray}

To see the uniqueness of $\lambda$ which makes diagram (\ref{eq:mulambdanu}) commute,
let $\xymatrix{\phi ! P \ar [r] ^{\lambda'} &Q}$ be another one as such.
First, note that, for any $V \in \Vbold$,
the naturality condition applied to $V \hookrightarrow \phi(C)$ gives that,
for every $[p,C]_{V} \in \phi ! P(V)$,
\begin{equation}
\lambda'_{V}([p,C]_{V}) = \lambda'_{\phi(C)}([p,C]_{\phi(C)})|_{V}.
\end{equation}
On the other hand,
since $\lambda'$ makes the diagram (\ref{eq:mulambdanu}) commutative,
we have 
\begin{equation}
\lambda'_{\phi(C)}([p,C]_{\phi(C)}) 
= 
\nu(p) 
=
\lambda_{\phi(C)}([p,C]_{\phi(C)}).
\end{equation}
Thus, it follows that
\begin{eqnarray}
\lambda'_{V}([p,C]_{V})
& = &
\lambda'_{\phi(C)}([p.C]_{\phi(C)})|_{V}\nonumber\\
& = &
 \lambda_{\phi(C)}([p.C]_{\phi(C)})|_{V}\nonumber\\
 & = &
 \lambda_{V}([p,C]_{V}).
\end{eqnarray}
Thus, it follows that $\lambda'_{V} = \lambda_{V}$, hence, $\lambda' = \lambda$.

\section{Lawvere-Tierney Topology, Grothendieck Topology, and Closure Operator}
\label{sec:Lawvere-Tierney Topology, Grothendieck}

In this appendix, we briefly review definitions of Lawvere-Tierney topologies, 
Grothendieck topologies, and closure operators related to the topologies.
Our explanation is restricted to the presheaf topos $\Set^{\Vboldop}$.

A Lawvere-Tierney topology is a morphism $\xymatrix{\Omega \ar [r] ^{j} & \Omega}$
satisfying the commutative diagrams
\begin{equation}
\xymatrix{
&& \,\Omega \; \\
1  \ar [rru] ^{\true} \ar [rrd] _{\true} && \\
&   & \, \Omega \ar [uu] _{j} \,  ,    \\
}
\label{eq:LawvereTierneycondition1}
\end{equation}
\begin{equation}
\xymatrix{
&& \,\Omega\;  \\
\Omega  \ar [rru] ^{j} \ar [rrd] _{j} && \\
&   &  \,\Omega \ar [uu] _{j}   \, ,    \\
}
\label{eq:LawvereTierneycondition2}
\end{equation}
and
\begin{equation}
\xymatrix{
\Omega \times \Omega \ar [rr] ^-{\wedge} \ar [dd]_{j \times j}   &&   \,  \Omega \; \ar [dd] ^{j}\\
&& \\
\Omega \times \Omega \ar [rr] _-{\wedge}  &&    \, \Omega \,. \\
}
\label{eq:LawvereTierneycondition3}
\end{equation}
Here, in diagram (\ref{eq:LawvereTierneycondition3}),
the morphism $\xymatrix{\Omega \times \Omega \ar [r] ^-{\wedge}  & \Omega }$
is defined by the $V$-component maps
$\wedge_{V}:\Omega(V) \times \Omega (V) \to \Omega(V)$ 
\begin{equation}
\wedge_{V}(\omega_{1}, \omega_{2}) := \omega_{1} \cap \omega_{2} \in \Omega(V).
\end{equation}

A Grothendieck topology $J$ on $\Vbold$ is 
a subobject of the subobject classifier $\Omega$
which satisfies, for each $V \in \Vbold$, the following conditions:
\begin{itemize}
\item[(i)] The maximum sieve $\tmath_{V}$ is in $J(V)$;
\item[(ii)] (transitivity axiom) if $\omega \in J(V)$ and $\omega' \in \Omega(V)$ such that $\omega'|_{V'} \in J(V')$ for all 
$V' \in \omega$, then $\omega' \in J(V)$.
\end{itemize}

The closure operator $\overline{(\,\cdot\,)}$ is a map which 
assigns, to each $S \in \Sub_{\Vboldhat}(Q)$,  $\overline{S} \in \Sub_{\Vboldhat}(Q)$
for every $Q \in \Set^{\Vboldop}$, in such a way as to satisfy,
for any $S$, $T \in \Sub_{\Vboldhat}(Q)$,
\begin{equation}
S \subseteq \overline{S},
\quad
\overline{\overline{S}} = \overline{S},
\quad
\mbox{and}
\quad
\overline{S \cap T} = \overline{S} \cap \overline{T}.
\label{eq:closurecondition}
\end{equation}

It is well known that the Lawvere-Tierney topology,
the Grothendieck topology, and the closure operator
are equivalent each other;
that is, each one of them induces the others uniquely.

Suppose that we are given a Lawvere-Tierney topology $j$.
Then, we can define the corresponding closure operator
from the following diagram:
\begin{equation}
\xymatrix{
\overline{S} \ar [rrrr] ^{!} \ar @{>->} [ddd] _{\overline{\iota}} &&&& \;1\; \ar @{>->} [ddl] _{\true} \ar @{>->} [ddd] ^{\true}\\
& S\ar [rrru] ^{!} \ar @{>->} [ddl] _{\iota} &&& \\
&&& \Omega \ar [rd] ^{j} \\
Q \ar [rrru] ^{\chi} \ar [rrrr]_{j \circ \chi} &&&& \;\Omega\,. \\
}
\label{eq:closure-j}
\end{equation}
Here, the inner square is a pullback.
Namely, the morphism $\chi$ is the characteristic morphism of
the subobject $\xymatrix{S \ar @{>->} [r] ^{\iota} & Q}$ of $Q$.
The closure $\overline{S}$ of $S$ is given as a subobject of $Q$
whose characteristic morphism equals $j \circ \chi$.
In other words,
it is defined in such a way that the outer square becomes a pullback.

Next, suppose that we have a closure operator $\overline{(\,\cdot\,)}$.
Then, we can obtain the corresponding Grothendieck topology $J$
as the closure of the terminal object $1$.
That is, regarding the terminal object $1(V) = \{\pt \} \simeq \{\tmath_{V}\}$
as a subpresheaf $\xymatrix{1 \ar @{^{(}->} @<-0.5mm>[r] ^{\true} & \Omega}$ of $ \Omega$,
we obtain the Grothendieck topology $J$ by
\begin{equation}
\xymatrix{ J \ar @{^{(}->} @<-0.5mm> [r] & \Omega} 
:= \xymatrix{\overline{1} \ar @{^{(}->} @<-0.5mm> [r] ^{\overline{\true}} & \Omega}.
\end{equation}

Finally, suppose that we have a Grothendieck topology $J$.
Then, the corresponding Lawvere-Tierney topology $j$ is defined 
as the characteristic morphism of $\xymatrix{ J \ar @{^{(}->} @<-0.5mm> [r] & \Omega} $.
Namely, $j$ is determined as a morphism making the diagram
\begin{equation}
\xymatrix{
J \ar [rr] ^{!} \ar @{^{(}->} @<-0.5mm> [dd] && 1 \ar @{^{(}->} @<-0.5mm> [dd] ^{\true} \\
&& \\
\Omega \ar [rr] _{j} && \Omega \\
}
\label{eq:J-j}
\end{equation}
a pullback.
If we start with a given $j$,
and in turn construct $\overline{(\,\cdot\,)}$ from $j$, $J$ from $\overline{(\,\cdot\,)}$, 
and $j'$ from $J$, in turn,
the the first $j$ and the final $j'$ are identical;
the Lawvere-Tierney topology, the Grothendieck topology,
and the closure operator mutually correspond bijectively.


\end{document}